\documentclass[reprint, aip, amsmath, amssymb, titlepage, 
          aps, pra, superscriptaddress, raggedbottom,
          nofootinbib, floatfix, citeautoscript]{revtex4-2}
\usepackage{silence}
\usepackage[utf8]{inputenc}
\usepackage[english]{babel}
\usepackage{hyperref}
\usepackage{bm}
\usepackage{hyperref}
\usepackage{float}
\usepackage[newcommands]{ragged2e}
\usepackage{physics}
\usepackage{tabularx}
\usepackage{mathtools}
\usepackage[raggedright]{titlesec}
\usepackage[toc,page,header]{appendix}
\usepackage{minitoc}
\usepackage{nicefrac}
\usepackage{sidecap}
\usepackage{multibib}
\usepackage[dvipsnames]{xcolor}
\usepackage{booktabs}
\usepackage{float}
\usepackage{graphicx}
\usepackage[capitalise]{cleveref}
\usepackage[version=4]{mhchem}
\usepackage{tabularx}
\usepackage{ulem}
\usepackage{enumitem}
\usepackage[per-mode=power, 
            exponent-product = \times, 
            inter-unit-product = \ensuremath{{\cdot}}, 
            tight-spacing = true,
            number-unit-product=~,
            parse-numbers=false,]{siunitx}
\AtBeginDocument{\RenewCommandCopy\qty\SI}
\ExplSyntaxOn
\msg_redirect_name:nnn { siunitx } { physics-pkg } { none }
\ExplSyntaxOff
\hypersetup{colorlinks=true,  linkcolor=blue, 
            citecolor=blue, filecolor=blue,  
            urlcolor=blue}
\usepackage{caption}
\titleformat*{\section}{\bfseries\large}
\titleformat*{\subsection}{\bfseries}
\titlespacing{\section}{0em}{1em}{0.25em}
\renewcommand{\textcite}[1]{\citenum{#1}}
\newcolumntype{Y}{>{\centering\arraybackslash}X}
\Crefformat{appendix}{Supplementary Information #2#1#3}
\let\origaddcontentsline\addcontentsline
\newcommand{\disabletocentries}{%
  \renewcommand{\addcontentsline}[3]{}%
}
\newcommand{\enabletocentries}{%
  \let\addcontentsline\origaddcontentsline
}

\begin{document}
\disabletocentries

\author{Pradyoth~Shandilya}
\affiliation{University of Maryland Baltimore County, Baltimore, MD, USA}
\author{Shao-Chien Ou}%
\affiliation{Joint Quantum Institute, NIST/University of Maryland, College Park, USA}
\affiliation{Microsystems and Nanotechnology Division, National Institute of Standards and Technology, Gaithersburg, USA}
\author{Alioune~Niang}
\affiliation{University of Maryland Baltimore County, Baltimore, MD, USA}
\author{Gary~Carter}
\affiliation{University of Maryland Baltimore County, Baltimore, MD, USA}
\author{Curtis~R.~Menyuk}
\affiliation{University of Maryland Baltimore County, Baltimore, MD, USA}
\author{Kartik~Srinivasan}
\affiliation{Joint Quantum Institute, NIST/University of Maryland, College Park, USA}
\affiliation{Microsystems and Nanotechnology Division, National Institute of Standards and Technology, Gaithersburg, USA}
\author{Gr\'egory~Moille}%
\email{gmoille@umd.edu}
\affiliation{Joint Quantum Institute, NIST/University of Maryland, College Park, USA}
\affiliation{Microsystems and Nanotechnology Division, National Institute of Standards and Technology, Gaithersburg, USA}
\date{\today}

\newcommand{\mytitle}{Universal Bright-Bright Integrated Soliton Molecule via Parametric Binding}
\title{\mytitle}
\begin{abstract}
Dissipative Kerr solitons (DKSs) have emerged as the preferred solution for on-chip integrated optical frequency comb (OFC) generation in metrology. %
A multi-pumped DKS enables either all-optical trapping in the Kerr-induced synchronization regime, or a multi-component OFC with a locked repetition rate yet with constant frequency offsets between the components in the multi-color DKS regime. %
The multi-color DKS regime is of particular interest since nonlinear mixing between the DKS and the secondary pumped component generates idler waves at different frequencies that are useful for spectral extension of the DKS comb. %
Here, we explore multi-color idler generation at frequencies in which the resonator free spectral range matches that at the DKS. %
We demonstrate theoretically and experimentally that without phase matching, the idler forms a bright pulse fundamentally bound to the bright DKS through parametric interaction, despite occurring in normal dispersion. %
Our work can enable new applications in metrology and spectroscopy of quantum systems toward visible wavelengths, as the parametric nature of our bright-bright state eliminates dependence on dispersion regime or visible wavelength pumping. %
\end{abstract}
\maketitle

\section{Introduction}
Time and frequency metrology has been revolutionized by ultra-precise phase control of optical pulse trains through stabilized octave-spanning optical frequency combs (OFCs). %
Such discovery has enabled ultra-precise atomic~\cite{OelkerNat.Photonics2019} and molecular~\cite{ZhangNature2024} optical clocks, exoplanet detection~\cite{MetcalfOpticaOPTICA2019a}, ultra-low noise microwave generation~\cite{FortierNaturePhoton2011}, and optical frequency synthesis~\cite{HolzwarthPhys.Rev.Lett.2000}. %
Meanwhile, advances in integrated photonics with enhanced capabilities as high-finesse cavities now achieve efficient nonlinear interactions with low input powers that are compatible with on-chip lasers~\cite{MoodyAVSQuantumSci.2020}. This has enabled the creation of on-chip integrated OFCs~\cite{SternNature2018a,ZhangNature2019,WangNature2024} compatible with low-cost mass-scale fabrication~\cite{LiuNatCommun2021a,OuOpt.Lett.OL2025}. %
A preferred OFC generation scheme leverages the $\chi^{3}$ nonlinear interaction to generate a dissipative Kerr soliton (DKS) under appropriate continuous-wave pumping~\cite{KippenbergScience2018} and offering low size, weight and power. %
This single pulse circulates in a microring resonator and is periodically extracted in the bus waveguide, creating a pulse train that forms the integrated OFC. %
Photonic engineering via microring dimension tuning tailors chromatic dispersion of the group velocity~\cite{OkawachiOpt.Lett.2014} for octave-spanning operation~\cite{LiOpticaOPTICA2017} and carrier-envelope offset detection and locking~\cite{SpencerNature2018,NewmanOptica2019}. %
This DKS stabilization is greatly simplified through all-optical Kerr-induced synchronization (KIS) with a secondary laser capturing one comb tooth~\cite{MoilleNature2023,MoilleOptica2025}, enabling orthogonal dual-point locking of the captured tooth and the carrier envelope offset~\cite{MoilleNature2023,MoilleOptica2025} or two comb teeth~\cite{SunNat.Photon.2025,Diakonov2025}, facilitating optical frequency division for a clockwork. %

Although KIS is of interest for metrology applications, dual-driven DKS beyond synchronization enables multi-color DKS~\cite{MenyukOpt.ExpressOE2025}, offering new operational regimes and enhanced comb performance. %
In a multi-color DKS, the DKS and secondary pumped component share identical group velocity through cross-phase modulation binding, but with different phase velocities. %
The phase offset accumulating each round trip produces two comb components with equal teeth separation but different offsets. %
This property enables distinct phase matching for DKS and secondary pump fields, creating synthetic dispersive waves DWs, that is, DWs that occur at locations that are not purely defined by the resonator dispersion, but instead through a combination of resonator dispersion and choice of secondary pump frequency~\cite{ZhangNatCommun2020}. %
These fields can nonlinearly interact to generate an idler at a different azimuthal modes with equal but opposite phase offset, significantly extending the frequency comb spectrum~\cite{MoilleNat.Commun.2021}. %
A particularly interesting multi-DKS regime occurs when the two pumps operate where the resonator supports a matching free-spectral range (FSR), demonstrated to create bright-dark bound states~\cite{ZhangPhys.Rev.Lett.2022}. %
This regime is notable because the dark pulse forms solely through cross-phase modulation XPM with the DKS, without relying on the dual balance of loss/driving field and dispersion/nonlinearity, enabling for much greater flexibility in the comb design and conversion efficiency. %
Yet, a comprehensive explanation of this effect remains elusive. %
Additionally, if the FSR-matched secondary pump creates these interesting regimes, we must ask if one could parametrically create such bound states through the idler, enabling access to otherwise unreachable wavelengths, and what the nature of such hypothetical bound states would be.

In this work, we present a comprehensive demonstration of multi-color bound states in a FSR-matched, multi-pumped DKS microcomb. %
We demonstrate two fundamentally different states created when either the secondary pump or idler are FSR matched. %
The former, corresponding to bound dark pulse formation is driven mainly by secondary pump depletion while the latter arises purely from parametric interaction, and frequency translates the soliton bright pulse bound to the DKS. %
Unlike previously reported dual-pumped soliton molecules where independent pulses couple,\cite{WengNatCommun2020a,LucasPhys.Rev.Lett.2025} this represents a parametric master-slave binding where the idler is a replica of the DKS. %
Importantly, the idler's bright pulse exists regardless of dispersion regime, allowing formation even under normal dispersion.
We demonstrate this bright-bright DKS-idler bound state through multi-color soliton simulations that accurately reproduce our experimental results in an integrated octave-spanning microring resonator, with the idler bright pulse forming at a near-visible wavelength. %

\section{Results}
\subsection{Concept}

\begin{figure}[!t]
     \centering
     \includegraphics[width=\columnwidth]{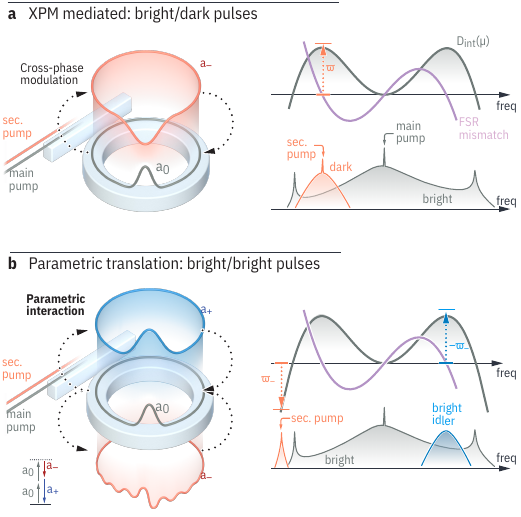}
     \caption{\label{fig:1} Concept of multi-pumped soliton bound states induced by nonlinearity in microring resonators. %
     \textbf{a} When dual pumping a microring resonator with the main pump generating the soliton and the secondary pump set outside phase matching---occurring when the phase offset $\varpi$ between the DKS and secondary pumped field exceeds the maximum integrated dispersion $D_\mathrm{int}(\mu)$---a bright-DKS/dark-pulse bound state forms. %
          The dark pulses arise from the cross-phase modulation (XPM) frequency shift experienced by the secondary pumped field when the DKS amplitude along the azimuthal coordinate is significant. %
     \textbf{b} The secondary pump field and the soliton can parametrically nonlinearly interact and generate an idler field at equal but opposite phase shift $-\varpi$. %
     Similar to the previous case, if $-\varpi$ is set to prevent phase matching, a bound state idler pulse with the DKS occurs. %
     The idler parametric driving field closely follows the soliton profile, hence only creating new light at the idler for sufficient energy in the soliton profile, resulting in a bright pulse regardless of the dispersion regime at the idler central mode. 
     }
\end{figure}

First, we aim to describe how a multi-pumped soliton creates bound-pulsed states. %
Two distinct cases are to be considered. %
Zhang et al~\cite{ZhangPhys.Rev.Lett.2022} reported the first: a main pump generates a bright DKS while a secondary pump generates a dark pulse if correctly set. %
Although they demonstrate this state through simulations and experiments, a theoretical explanation remains absent. %
The second case, our focus, results in the bright pulse generation at a central frequency different from both pumps through solitonic parametric interaction. 
To understand both cases, we use the multi-color soliton framework~\cite{MenyukOpt.ExpressOE2025}. %
The starting point is the modified Lugiato-Lefever equation (mLLE) accounting for multiple driving forces~\cite{TaheriEur.Phys.J.D2017}: 

\begin{equation}
     \label{eq:mLLE}
     \begin{split}
          \frac{\partial a(\theta,t)}{\partial t} =& -\left( \frac{\kappa}{2} + i \delta\omega \right)a + i \mathcal{D}(a) -2i\gamma |a|^2a \\
          &+ F_0  + F_- \mathrm{exp}(i\varpi_- t + i \mu_- \theta)
     \end{split}
\end{equation}

\noindent with $\varpi_- = \delta \omega_\mathrm{-} - D_\mathrm{int}(\mu_-)$ the frequency offset of the secondary pump to its closest DKS microcomb tooth at mode $\mu_-$ ($\mu_-$ is indexed relative to the main pump mode $\mu=0$), and $D_\mathrm{int}(\mu) = \omega_\mathrm{res}(\mu) - \left( \omega_0 + \omega_\mathrm{rep}\mu \right)$ with $\omega_0$ the main pump frequency, and $\omega_\mathrm{res}\left( \mu \right)$ the resonant frequency at mode $\mu$, $\mathcal{D}(a) = \sum_\mu D_\mathrm{int}(\mu)A(\mu, t)\mathrm{exp}^{i\mu\theta}$ the dispersion operator with $A(\mu, t)$ the azimuthal Fourier transform of $a(\theta, t)$, $\kappa\approx~$\qty{300}{\MHz} the total losses, $\gamma$ the effective nonlinear coefficient, $F_0=\sqrt{\kappa_\mathrm{ext}P_0}$ the main driving field with on-chip pump power $P_0$ and $F_-=\sqrt{\kappa_\mathrm{ext}P_-}$ the secondary driving field with on-chip pump power $P_-$, and $\kappa_\mathrm{ext} = \kappa/2$ the external coupling rate set to be critically coupled. %
We assume a weak group-velocity-dispersion over the frequency span of interest, allowing the Kerr nonlinearity to impose a single angular group velocity (and thus a single repetition rate) on all components. %
However, we consider the regime outside Kerr-induced synchronization~\cite{MoilleNature2023}, so that the nonlinearity is too slow for phase locking, resulting in a phase slip between different components of the overall nonlinear state, which we refer to as colors. %
This allows a decomposition of the field into the colors as $a = a_0  + a_- \mathrm{exp}\left(  i\varpi t\right) + a_+ \mathrm{exp}\left(  -i\varpi t\right)$, which when substituted into~\cref{eq:mLLE} yields:

\begin{subequations}
     \begin{align}
          \frac{\partial a_-(\theta,t)}{\partial t} =&-\left(\frac{\kappa}{2} + i \varpi_-\right) a_- +i\mathcal{D}(a_-)\nonumber\\
               &- i\gamma\left(|a_-|^2 + 2|a_0|^2 + 2|a_+|^2 \right)a_- \label{eq:signal}\\
               &- i\gamma a_0^2 a_+^* +  F_-\nonumber\\
          \frac{\partial a_0(\theta,t)}{\partial t} =&-\left(\frac{\kappa}{2} + i \delta\omega\right) a_0 +i\mathcal{D}(a_0)\nonumber\\
               &- i\gamma\left(2|a_-|^2 + |a_0|^2 + 2|a_+|^2 \right)a_0 \label{eq:dks}\\
               &- i\gamma a_0^* a_-a_+ +  F_0\nonumber\\
          \frac{\partial a_+(\theta,t)}{\partial t} =&-\left(\frac{\kappa}{2} - i \varpi_-\right) a_+ +i\mathcal{D}(a_+)\nonumber\\
               &- i\gamma\left(2|a_-|^2 + 2|a_0|^2 + |a_+|^2 \right)a_+ \label{eq:idler}\\
               &- i\gamma a_0^2 a_-^*\nonumber
     \end{align}
\end{subequations}

Here, $a_0$ is the main DKS, $a_-$ is referred to as the signal, and $a_+$ is referred to as the idler, in analogy to four-wave mixing interactions between continuous wave fields. %
To understand the bright-dark bound state, we focus on~\cref{eq:signal} assuming $a_-$ remains small compared to $a_0$ and that idler generation ($a_+$) is negligible. This yields:

\begin{equation}
     \label{eq:signal_approx}
     \begin{split}
          \frac{\partial a_-(\theta,t)}{\partial t} =&-\left[\frac{\kappa}{2} + i \Big( \varpi_- + 2\gamma|a_0(\theta,t)|^2 \Big)\right]a_-\\
               & + i\mathcal{D}(a_-) +F_-
     \end{split}
\end{equation}

If the secondary pump $F_-$ is set such that $\varpi_- \geq D_\mathrm{int}(\mu) \; \forall \; \mu$ [\cref{fig:1}a], $a_-$ will not experience any phase matched condition and no dispersive wave will be generated. %
In such a case, the azimuthal angle dependent cross-phase modulation (XPM) from the interaction with the soliton $a_0$ creates, where $|a_0(\theta, t)|^2$ is significant, a frequency shift that further detunes $a_-$ from any phase matching condition.
This results in a depletion of $a_-$, bound to the soliton hyperbolic secant profile, creating an effective dark pulse with its width dependent on the phase-mismatch with the closest azimuthal mode, hence the second order dispersion term around $\mu_-$.

The second regime that we unveil in this manuscript is the creation of a bright-bright state due to the nonlinear interaction between the soliton and the secondary pump, regardless of the dispersion at the idler central mode. %
To understand this effect, we can consider a similar approximation than made previously, that $a_+$ is small compared to $a_0$, which results in the following approximation of~\cref{eq:idler}:
\begin{equation}
     \begin{split}
          \frac{\partial a_+(\theta,t)}{\partial t} =&-\left[\frac{\kappa}{2} + i \Big( -\varpi_- + 2\gamma|a_0|^2 \Big)\right]a_+ \\
               &+ i\mathcal{D}(a_+) -  i\gamma a_0^2(\theta,t) a_-^*(\theta,t)
     \end{split}
\end{equation}

Similarly, we assume no phase matching for $a_+$, setting the system so that $-\varpi_->D_\mathrm{int}(\mu) \;\forall \mu$ to avoid any dispersive wave generation [\cref{fig:1}b] (\textit{i.e.} the signal field $a_-$ is at least phase matched at its pumped mode $\mu_-$). %
The same XPM term than in~\cref{eq:signal_approx} is present in the idler field, which creates a soliton profile dependent frequency shift. %
However, such an XPM term can be for the moment neglected in terms of its impact on the dynamics, as the idler driving term is purely parametric in origin and arises from the interplay of the signal $a_-$ (dominated by the CW pump) and the soliton field $a_0\left( \theta,t \right)$. 
The resulting driving field is negligible throughout the soliton pedestal, resulting in no idler generation, while being significant within the soliton pulse profile. 
Thus, this parametric interaction then always lead to a bright pulse, regardless of the sign of the dispersion at the idler mode $\mu_+$. %
Importantly, regardless of the phase-(mis)matching condition of $a_\pm$, the Kerr nonlinearity binds their group velocity to $a_0$, a core hypothesis for the multi-color decomposition of~\cref{eq:signal,eq:dks,eq:idler}.

Interestingly, by deriving the solution in the azimuthal space (see methods) the idler field $A_+(\mu_,t)$ is expressed: 
\begin{equation}
     \label{eq:idler_approx}
      \begin{split}
        A_+\left(\mu-\mu_+\right) &= \frac{2\gamma}{i\frac{\kappa}{2}  + \frac{D^{(+)}_2}{2} (\mu-\mu_+)^2 } F_0F_-^*A_0(\mu,t) \\
    \end{split}
\end{equation}

Which corresponds to spectral translation of the DKS azimuthal component $A_0(\mu,t)$ (the main soliton) to the region around mode $\mu_+$ through four-wave mixing Bragg scattering~\cite{LiNaturePhoton2016,QureshiCommunPhys2022} from the moving grating induced by $F_0$ and $F_-^*$, hence preserving all DKS properties, including its bright pulse nature. %
Critically, the idler is not an independent DKS. %
While the parametric driving term $-i\gamma a_0^2(\theta,t) a_-^*(\theta,t)$ can in principle generate solitonic structures~\cite{Weng2025}, here it acts as a spectral translation mechanism, creating a slave field that replicates the main DKS. %
The idler does not satisfy the dual balance of dispersion/nonlinearity and gain/loss characteristic of an independent DKS; rather, it is a parametric replica of the master soliton, hence a bright pulse. %
This parametric master-slave binding distinguishes it from molecules formed by coupling between independent solitons. %
Additionally, since the idler inherits the DKS properties, in the case of a dark DKS pulse (i.e., the main pump in the normal dispersion regime), the idler will be a dark pulse regardless of the dispersion at its center mode.

\begin{figure*}[t]
     \centering
     \includegraphics{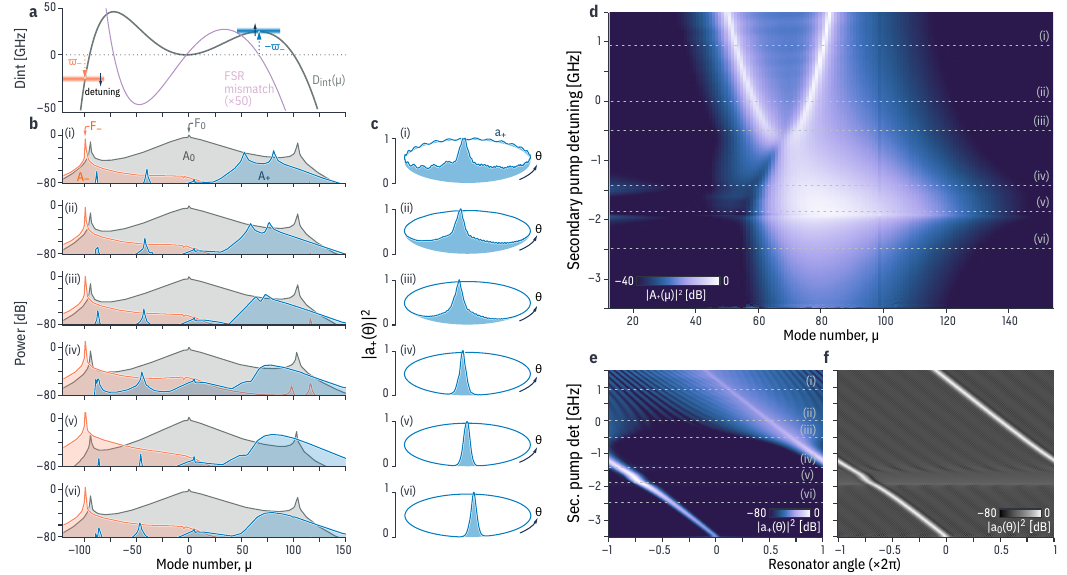}
     \caption{\label{fig:2}
     Numerical demonstration of parametric bright pulse formation through secondary pump detuning. %
     \textbf{a} Integrated dispersion profile $D_{\text{int}}(\mu)$ with main pump $F_0$ creating the DKS comb $A_0$ and the secondary pump $F_-$. %
     The integrated dispersion and FSR mismatch curve are computed directly from the finite element method simulated resonance frequencies. %
     The phase offset $\varpi_-$ of the secondary pumped comb with the DKS enables nonlinear phase mixing, creating an idler at an offset $-\varpi_-$ from the DKS comb. %
     The shaded area corresponds to the detuning used in the simulation. %
     \textbf{b} Spectral evolution of the DKS ($A_0$), secondary pump ($A_-$), and idler comb ($A_+$) with the secondary pump detuning from positive (phase-matched idler) to negative (phase-mismatched idler) values across six conditions (i-vi). 
     \textbf{c} Corresponding azimuthal profile of the idler color $|a_+(\theta)|^2$, transforming into a bright $\sech$ pulse as phase matching is prevented. %
     \textbf{d} Idler spectrum showing transition from synthetic dispersive waves at positive detuning to broadband bright pulse at negative detuning. %
     \textbf{e-f} Azimuthal profiles comparing idler intensity of the idler color $|a_+(\theta)|^2$ (e) with the DKS $|a_0(\theta)|^2$ (f), demonstrating that the parametric idler forms a bright pulse co-localized with the soliton when phase matching is prevented. %
     }
\end{figure*}

\begin{figure*}[t]
     \centering
     \includegraphics{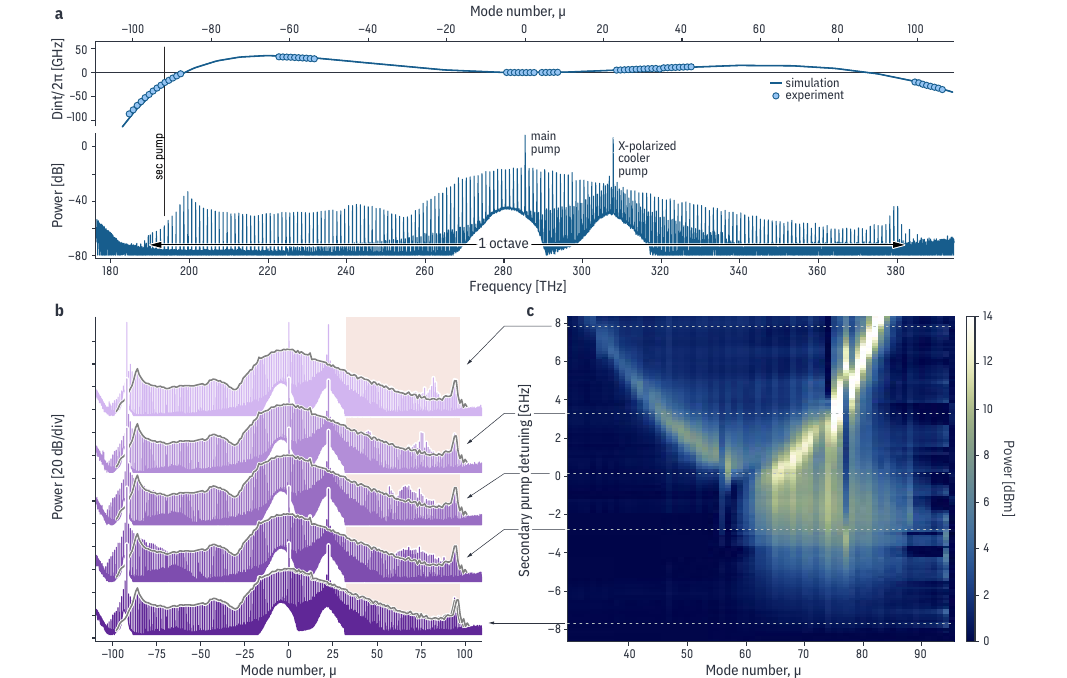}
     \caption{\label{fig:3}
      Experimental demonstration of the bright-bright parametrically binded pulse regime. 
      \textbf{a} We use an integrated microring resonator designed to support octave-spanning operation of the microcomb with an integrated dispersion $D_\mathrm{int}(\mu)$ (top) presenting two zero-crossing at \qty{199}{\THz} and \qty{378}{\THz} when assuming a pump at \qty{285.3}{\THz}. %
      The resulting single-pump DKS, obtained with thermal-stabilization using a cross-polarized cooler pump at \qty{307.7}{\THz} present two dispersive waves at the expected phase-matched mode from $D_\mathrm{int}$ and spanning over an octave. %
      \textbf{b} We then introduce a secondary pump at \qty{193.5}{\THz} to explore bright-bright DKS-idler bond state formation. %
      As the theory predicts, the experimental idler comb profile changes with secondary pump detuning based on phase matching presence or absence. %
      \textbf{c} We reproduce the idler comb map experimentally, with striking resemblance to the theoretical finding of \cref{fig:2}, highlighting the parametric bright pulse formation when the phase matching is prevented. %
     }
\end{figure*}

\begin{figure}[t]
     \centering
     \includegraphics{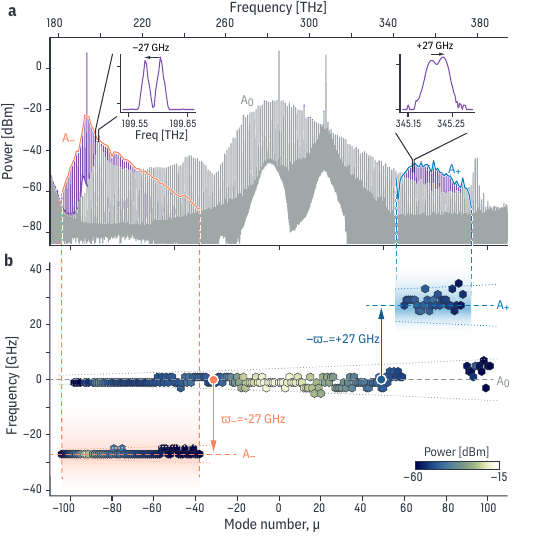}
     \caption{\label{fig:4}
     Experimental demonstration of repetition rate binding. %
     \textbf{a} The measured spectrum in the multi-color soliton regime exhibits three distinct comb components equally offset from one another with opposite signs, confirming multi-color soliton theory and phase conservation. %
     \textbf{b} From the spectrum, one extracts the relative offset of each comb tooth against a fixed frequency grid. %
     The DKS matches the grid perfectly, while the secondary pump and idler comb teeth show equal but opposite frequency offsets, as theory predicts. %
     Both exhibit zero slope (dashed lines), indicating identical repetition rates to the DKS, confirming the binding of all three colors. The dotted lines represent the OSA resolution error which is frequency dependent. %
     }
\end{figure}

\subsection{Numerical investigation}

To demonstrate this new bright/bright binded regime, we simulate the multi-color soliton of~\cref{eq:signal,eq:dks,eq:idler} dynamically using the open-source pyLLE package~\cite{MoilleJ.RES.NATL.INST.STAN.2019}. %
Starting from numerical noise, we sweep the main pump detuning to generate a stable DKS, then introduce the secondary pump, which with a parked main pump, we only sweep the secondary pump frequency detuning to explore the phase-matched and phase-mismatched regimes. %
We consider the integrated dispersion $D_\mathrm{int}(\mu) = \omega_\mathrm{res}(\mu) - (\omega_\mathrm{0} + \mu\omega_\mathrm{rep})$ that presents two zero crossings~[\cref{fig:2}a]. %
Here, $\omega_\mathrm{res}(\mu)$ is the angular frequency of resonance for mode $\mu$ normalized to the pumped mode, $\omega_\mathrm{0}$ is the main pump frequency at mode $\mu=0$, and $\omega_\mathrm{rep}$ is the repetition rate of the DKS. %
Although~\cref{eq:idler_approx} requires only that $D_\mathrm{int}$ has a local maximum at mode $\mu_+$ (ensuring one zero crossing from the $D_\mathrm{int}(\mu)$ roll-off), corresponding to a region where the FSR matches that at the main pump, we selected this dispersion to enable on-resonance operation of the secondary pump at $\mu_-=-99$, exhibiting a comb tooth frequency mismatch $-\omega_- \approx \max(D_\mathrm{int}(\mu>0))$ allowing for the secondary pump detuning alone to switch between the phase-matched and phase-mismatched regimes of the idler color. %
For positive secondary pump detuning [\cref{fig:2}b, conditions (i-ii)], we enable the phase-matching condition such that $-\varpi_- = D_\mathrm{int}(\mu>0)$ exists at two distinct azimuthal modes. %
The idler spectrum [\cref{fig:2}b-d] exhibits two prominent peaks corresponding to these phase-matched modes, while the azimuthal profile [\cref{fig:2}c] remains relatively flat, characteristic of continuous-wave-like behavior with minimal structure. %
As the frequency detuning is further red tuned [\cref{fig:2}b,conditions (iii-iv)], the frequency shifting of the idler is increased further away from the DKS, modifying the phase-matched modes according to the $D_\mathrm{int}(\mu)$ spectral profile. %
The transition occurs when the secondary pump detuning becomes sufficiently negative that $-\varpi_- > D_\mathrm{int}(\mu)$ $\forall \mu>0$ [\cref{fig:2}b-c conditions (v-vi)], eliminating all phase matching conditions. %
The idler spectrum [\cref{fig:2}b-d] transforms from the discrete synthetic dispersive wave peaks to a smooth, broadband $\mathrm{sech}^2$ envelope characteristic of a bright pulse. %
Correspondingly, the azimuthal profile [\cref{fig:2}c] evolves into a localized bright pulse that precisely overlaps with the DKS position [\cref{fig:2}e-f]. %
Importantly, the idler forms this bright pulse despite the normal dispersion at its central mode $\mu_+$, confirming that the parametric driving mechanism described by~\cref{eq:idler_approx} dominates over the dispersive effects and XPM-mediated secondary pump depletion. %
The parametric nature of the interaction ensures that idler generation occurs only where the product $a_0^2 a_-^*$ is significant, which restricts generation to the soliton pulse region and naturally produces a bright pulse independent of the local dispersion regime. %
Although the idler bright pulse appears to sit on a zero background, it is in fact non-null. %
The parametric driving term $-i\gamma a_0^2 a_-^*$ involves both the DKS field $a_0$, which sits on a pedestal with its main spectral component at the pump mode $\mu=0$, and the secondary pump field $a_-$, which is predominantly a CW wave filling the cavity azimuthally. %
The parametric interaction between these CW tones is highly phase mismatched due to the asymmetry of the system $\mu_+ \neq -\mu_-$, leading to a very weak pedestal for the idler pulse, yet non-zero. %
Another interesting observation is the presence of dark spectral modes [\cref{fig:2}d], which drift linearly with the secondary pump detuning and are reproduced experimentally [\cref{fig:3}c]. %
We hypothesize this may arise from XPM-induced envelope distortion, similar to observations in bright-dark molecules~\cite{ZhangPhys.Rev.Lett.2022}, or from nonlinear PT-symmetry effects~\cite{MiriNewJ.Phys.2016,ParkPhys.Rev.Lett.2021}, though a definitive explanation remains an object of ongoing investigation.

\subsection{Experimental demonstration}

Experimentally, we demonstrate the generation of the parametric bright-bright soliton bound state in an integrated silicon nitride (SiN) microring resonator embedded in silicon dioxide. %
The microring has SiN thickness $H=\qty{650}{\nm}$, radius $RR=\qty{23}{\um}$, and width $RW=\qty{850}{\nm}$, enabling tailored group velocity dispersion for octave-spanning operation from near-infrared to the edge of the visible. %
The bus-to-ring coupling uses a pulley configuration with a \qty{460}{\nm} wide waveguide, \qty{19}{\um} coupling length, and \qty{500}{\nm} gap. %
This achieves critical coupling at the pump wavelength ($Q_\mathrm{tot}\approx0.75\times10^6$), overcoupling in the C-band ($Q_\mathrm{tot}\approx0.25\times10^6$), and efficient extraction at short wavelengths.%

The resonator's integrated dispersion shows excellent agreement between finite element simulations and the calibrated measurements, with phase-matched modes at \qty{199}{\THz} and \qty{378}{\THz}~[\cref{fig:3}a]. %
The DKS is generated by pumping the cavity with a \qty{150}{\mW} on-chip power main pump laser at \qty{285.3}{\THz} [\cref{fig:3}b]. %
The resonator is thermally stabilized with a \qty{307.7}{\THz} cooler laser~\cite{ZhangOptica2019,ZhouLightSciAppl2019}  in a cross-polarized, co-propagating configuration with the main pump. %
The secondary pump $F_-$, with on-chip power \qty{P_-=50}{\mW}, is set at azimuthal mode $\mu_-=-92$ (\qty{193.58}{\THz}) below the low-frequency dispersive wave $\mu_\mathrm{DW}=-86$ (\qty{199.6}{\THz}), enabling on-resonance $\varpi_- < 0$ operation for efficient idler generation around the $D_\mathrm{int}(\mu)$ maximum at $\mu_+=61$. %
We use a standard fiber-optics approach for wavelength multiplexing and on-chip light delivery, with further details on the experimental apparatus available in our previous work~\cite{MoilleNat.Commun.2021,MoilleNat.Photon.2024a}. %
We tune the secondary laser frequency over \qty{16}{\GHz} using its piezo-electric element, controlling phase (mis)matching of the idler to  switch between the dispersive wave and parametric bright pulse regime. %
For positive detuning of the secondary pump, the idler remains phase-matched with azimuthal modes, generating two synthetic DWs [\cref{fig:3}b]. %
Further detuning toward negative frequency leads to a sech envelope that can be observed in-between the main pump and the high frequency dispersive wave at $\mu=75$, and is the signature of the parametric bright pulse generation. %
Since the DKS comb spectrum is identical in the single and multi-color soliton regimes, we extract the idler profile by subtracting the single-pump spectrum from the multi-pump spectrum at each secondary pump detuning step of \qty{\approx 150}{MHz} [\cref{fig:3}c]. %
The experimental results match the numerical simulation presented in \cref{fig:2} very closely. %
The synthetic DW follows the integrated dispersion profile, while the parametric bright pulse emerges once no phase matching condition is met. %
Importantly, we reproduce the numerical results where a few dark modes that are anti-crossing with the bright pulse are observed and result from the nonlinear interaction between the three colors rather than the direct linear avoided mode crossing(s) observed in such cases in both the synthetic DW and bright parametric pulse regimes. %
Additionally, we also observe that the idler pulse central mode is not at the mode where $D_\mathrm{int}(\mu)$ is maximum, but rather is offset toward higher $\mu$.

Finally, we verify experimentally the bright-bright state bonding~[\cref{fig:4}], which follows from theory since the idler forms only when the soliton energy is significant and forces coexistence at the same azimuthal angle. %
We measure the full optical spectrum in the multi-color regime~[\cref{fig:4}a], which exhibits the three distinct comb components: the DKS at its native carrier-envelope offset frequency, and the secondary pump and idler colors shifted from the DKS microcomb. %
From such spectra, we extract the spectral component of each $\mu$ mode. %
Centering each $\mu$ around a DKS tooth, we extract the spectral power within $\left[-\omega_\mathrm{rep}/2; \omega_\mathrm{rep}/2\right]$, corresponding to a full $2\pi$ phase offset from the DKS. %
A repetition rate discrepancy between the DKS and other colors appears as a slope in this representation, while a phase velocity discrepancy alone creates a fixed offset from the DKS grid. %
We observe that the DKS comb teeth align perfectly with the reference grid (zero offset), confirming our analysis method, while the secondary pump and idler comb teeth exhibit constant frequency offsets of $\approx \mp\qty{27}{\GHz}$, respectively, with no slope across more than twenty modes [\cref{fig:4}b]. %
All three colors exhibit zero slope across their respective mode ranges (dashed lines), confirming identical repetition rates, with the secondary pump and idler colors offset equally but oppositely from the DKS. %
This validates the cross-phase modulation binding mechanism that imposes a single group velocity on all components while allowing phase velocity differences, and confirms that the parametric idler is bound to the DKS through the nonlinear interaction described by~\cref{eq:signal_approx,eq:idler_approx}. %

\section{Discussion}
We have provided a comprehensive theoretical framework and experimental demonstration of multi-color bound states in FSR-matched multi-pumped dissipative Kerr solitons. While the bright-dark molecule was previously observed experimentally,\cite{ZhangPhys.Rev.Lett.2022} we provide the first theoretical explanation of its formation through XPM-induced pump depletion when $\varpi_- \geq D_{\text{int}}(\mu)$ $\forall\mu$ prevents phase matching. %
Building on this framework, we introduce and demonstrate the bright-bright molecule, where the bright idler pulse emerges purely from parametric interaction between the DKS and secondary pump, translating the soliton envelope to a new spectral region through four-wave mixing Bragg scattering\cite{MoilleNat.Commun.2021}. \\%
\indent The parametric bright pulse forms regardless of dispersion regime at the idler wavelength. %
By preventing phase matching ($-\varpi_- > D_{\text{int}}(\mu)$ $\forall\mu$), the idler generates only at an azimuthal angle where the DKS is intense, producing a bright pulse even in normal dispersion. %
This decouples bright pulse generation from dispersion engineering, enabling access to visible wavelengths without the extreme geometry scaling that approaches fabrication limits and degrades quality factors in silicon nitride platforms~\cite{MoilleNature2023}. %
The main DKS operates in the near-infrared where anomalous dispersion is readily achievable, while the parametric idler extends the comb spectrum to near-visible atomic transitions such as rubidium (780~nm) and cesium (852~nm). %
Moreover, the bright pulse has minimal spectral background compared to dispersive waves, reducing filtering requirements and improving coupling to atomic systems, which is particularly valuable for optical atomic clocks, quantum information processing, and precision spectroscopy requiring clean spectral lines at specific atomic transitions. %
Although in this work we use a standard rectangular cross-section microring resonator, Fourier synthesis dispersion engineering via photonic-crystal corrugations~\cite{MoilleCommunPhys2023} offers a potential improvement to finely control the creation of new local maxima in the integrated dispersion where bright idler pulses can be generated, expanding the idler's accessible wavelength range to address application-specific requirements. \\%
\indent The multi-color framework enables flexible stabilization pathways. %
Building on our Kerr-induced synchronization work,\cite{MoilleNature2023,MoilleOptica2025} parametric binding enables indirect stabilization: all-optical trapping of the main DKS through KIS~\cite{MoilleNature2023} stabilizes all other colors via optical frequency division~\cite{MoilleNat.Photon.2024a}. %
This stabilizes the entire system, including the idler color, through nonlinear coupling even when direct laser access or stabilization at the idler wavelength is impractical. \\%
This substantially expands the operational parameter space for integrated optical clocks and frequency synthesizers as the secondary pump power and detuning provide dynamic control over the idler spectrum, offering new freedom in ultra-stable Kerr comb engineering for applications requiring tailored spectral coverage at specific wavelengths. %

\section*{Acknowledgments}
P.S., A.N, G.C, and C.M. acknowledge the support from the Air Force Office of Scientific Research (Grant No. FA9550-20-1-0357), a collaborative agreement with the National Center for Manufacturing Sciences (contract nos. 2022138-142232 and 2023200-142386) as  sub-awards from the US Department of Defense (cooperative agreement nos. HQ0034-20-2-0007 and HQ0034-24-2-0001). 
P.S. and C.M. acknowledge the support from NIST (grant no. 60NANB24D106). %
A.N and G.C acknowledge  the support from the Army Research Office (contract no. W911NF- 22-S-0010). %
S.O, K.S., and G.M. acknowledge the partial funding support from the Space Vehicles Directorate of the Air Force Research Laboratory and the NIST-on-a-chip program of the National Institute of Standards and Technology. %

\section*{Author Declarations}

\subsection*{Conflict of Interest}
The authors have no conflicts to disclose.

\subsection*{Author Contributions}
P.S. developed the theory and performed the simulations with the help of C.M. and G.M.. %
G.M. performed the experimental work and led the project, with the help of K.S..  %
S.C. and A.N. helped with the early characterization of the devices, with the assistance of G.C.
G.M. wrote the manuscript with contributions from all authors.
All the authors contributed to and discussed the content of this article.

\subsection*{Data Availability}
The data that support the plots within this paper and other findings of this study are available from the corresponding author upon reasonable request.


\begin{thebibliography}{37}%
\makeatletter
\providecommand \@ifxundefined [1]{%
 \@ifx{#1\undefined}
}%
\providecommand \@ifnum [1]{%
 \ifnum #1\expandafter \@firstoftwo
 \else \expandafter \@secondoftwo
 \fi
}%
\providecommand \@ifx [1]{%
 \ifx #1\expandafter \@firstoftwo
 \else \expandafter \@secondoftwo
 \fi
}%
\providecommand \natexlab [1]{#1}%
\providecommand \enquote  [1]{``#1''}%
\providecommand \bibnamefont  [1]{#1}%
\providecommand \bibfnamefont [1]{#1}%
\providecommand \citenamefont [1]{#1}%
\providecommand \href@noop [0]{\@secondoftwo}%
\providecommand \href [0]{\begingroup \@sanitize@url \@href}%
\providecommand \@href[1]{\@@startlink{#1}\@@href}%
\providecommand \@@href[1]{\endgroup#1\@@endlink}%
\providecommand \@sanitize@url [0]{\catcode `\\12\catcode `\$12\catcode `\&12\catcode `\#12\catcode `\^12\catcode `\_12\catcode `\%12\relax}%
\providecommand \@@startlink[1]{}%
\providecommand \@@endlink[0]{}%
\providecommand \url  [0]{\begingroup\@sanitize@url \@url }%
\providecommand \@url [1]{\endgroup\@href {#1}{\urlprefix }}%
\providecommand \urlprefix  [0]{URL }%
\providecommand \Eprint [0]{\href }%
\providecommand \doibase [0]{https://doi.org/}%
\providecommand \selectlanguage [0]{\@gobble}%
\providecommand \bibinfo  [0]{\@secondoftwo}%
\providecommand \bibfield  [0]{\@secondoftwo}%
\providecommand \translation [1]{[#1]}%
\providecommand \BibitemOpen [0]{}%
\providecommand \bibitemStop [0]{}%
\providecommand \bibitemNoStop [0]{.\EOS\space}%
\providecommand \EOS [0]{\spacefactor3000\relax}%
\providecommand \BibitemShut  [1]{\csname bibitem#1\endcsname}%
\let\auto@bib@innerbib\@empty
\bibitem [{\citenamefont {Oelker}\ \emph {et~al.}(2019)\citenamefont {Oelker}, \citenamefont {Hutson}, \citenamefont {Kennedy}, \citenamefont {Sonderhouse}, \citenamefont {Bothwell}, \citenamefont {Goban}, \citenamefont {Kedar}, \citenamefont {Sanner}, \citenamefont {Robinson}, \citenamefont {Marti}, \citenamefont {Matei}, \citenamefont {Legero}, \citenamefont {Giunta}, \citenamefont {Holzwarth}, \citenamefont {Riehle}, \citenamefont {Sterr},\ and\ \citenamefont {Ye}}]{OelkerNat.Photonics2019}%
  \BibitemOpen
  \bibfield  {author} {\bibinfo {author} {\bibfnamefont {E.}~\bibnamefont {Oelker}}, \bibinfo {author} {\bibfnamefont {R.~B.}\ \bibnamefont {Hutson}}, \bibinfo {author} {\bibfnamefont {C.~J.}\ \bibnamefont {Kennedy}}, \bibinfo {author} {\bibfnamefont {L.}~\bibnamefont {Sonderhouse}}, \bibinfo {author} {\bibfnamefont {T.}~\bibnamefont {Bothwell}}, \bibinfo {author} {\bibfnamefont {A.}~\bibnamefont {Goban}}, \bibinfo {author} {\bibfnamefont {D.}~\bibnamefont {Kedar}}, \bibinfo {author} {\bibfnamefont {C.}~\bibnamefont {Sanner}}, \bibinfo {author} {\bibfnamefont {J.~M.}\ \bibnamefont {Robinson}}, \bibinfo {author} {\bibfnamefont {G.~E.}\ \bibnamefont {Marti}}, \bibinfo {author} {\bibfnamefont {D.~G.}\ \bibnamefont {Matei}}, \bibinfo {author} {\bibfnamefont {T.}~\bibnamefont {Legero}}, \bibinfo {author} {\bibfnamefont {M.}~\bibnamefont {Giunta}}, \bibinfo {author} {\bibfnamefont {R.}~\bibnamefont {Holzwarth}}, \bibinfo {author} {\bibfnamefont {F.}~\bibnamefont {Riehle}}, \bibinfo {author} {\bibfnamefont {U.}~\bibnamefont {Sterr}},\ and\ \bibinfo {author} {\bibfnamefont {J.}~\bibnamefont {Ye}},\ }\bibfield  {title} {\enquote {\bibinfo {title} {Demonstration of 4.8 {\texttimes} 10-17 stability at 1 s for two independent optical clocks},}\ }\href {https://doi.org/10.1038/s41566-019-0493-4} {\bibfield  {journal} {\bibinfo  {journal} {Nature Photonics}\ }\textbf {\bibinfo {volume} {13}},\ \bibinfo {pages} {714--719} (\bibinfo {year} {2019})}\BibitemShut {NoStop}%
\bibitem [{\citenamefont {Zhang}\ \emph {et~al.}(2024)\citenamefont {Zhang}, \citenamefont {Ooi}, \citenamefont {Higgins}, \citenamefont {Doyle}, \citenamefont {{von der Wense}}, \citenamefont {Beeks}, \citenamefont {Leitner}, \citenamefont {Kazakov}, \citenamefont {Li}, \citenamefont {Thirolf}, \citenamefont {Schumm},\ and\ \citenamefont {Ye}}]{ZhangNature2024}%
  \BibitemOpen
  \bibfield  {author} {\bibinfo {author} {\bibfnamefont {C.}~\bibnamefont {Zhang}}, \bibinfo {author} {\bibfnamefont {T.}~\bibnamefont {Ooi}}, \bibinfo {author} {\bibfnamefont {J.~S.}\ \bibnamefont {Higgins}}, \bibinfo {author} {\bibfnamefont {J.~F.}\ \bibnamefont {Doyle}}, \bibinfo {author} {\bibfnamefont {L.}~\bibnamefont {{von der Wense}}}, \bibinfo {author} {\bibfnamefont {K.}~\bibnamefont {Beeks}}, \bibinfo {author} {\bibfnamefont {A.}~\bibnamefont {Leitner}}, \bibinfo {author} {\bibfnamefont {G.~A.}\ \bibnamefont {Kazakov}}, \bibinfo {author} {\bibfnamefont {P.}~\bibnamefont {Li}}, \bibinfo {author} {\bibfnamefont {P.~G.}\ \bibnamefont {Thirolf}}, \bibinfo {author} {\bibfnamefont {T.}~\bibnamefont {Schumm}},\ and\ \bibinfo {author} {\bibfnamefont {J.}~\bibnamefont {Ye}},\ }\bibfield  {title} {\enquote {\bibinfo {title} {Frequency ratio of the {{229mTh}} nuclear isomeric transition and the {{87Sr}} atomic clock},}\ }\href {https://doi.org/10.1038/s41586-024-07839-6} {\bibfield  {journal} {\bibinfo  {journal} {Nature}\ }\textbf {\bibinfo {volume} {633}},\ \bibinfo {pages} {63--70} (\bibinfo {year} {2024})}\BibitemShut {NoStop}%
\bibitem [{\citenamefont {Metcalf}\ \emph {et~al.}(2019)\citenamefont {Metcalf}, \citenamefont {Anderson}, \citenamefont {Bender}, \citenamefont {Blakeslee}, \citenamefont {Brand}, \citenamefont {Carlson}, \citenamefont {Cochran}, \citenamefont {Diddams}, \citenamefont {Endl}, \citenamefont {Fredrick}, \citenamefont {Halverson}, \citenamefont {Hickstein}, \citenamefont {Hearty}, \citenamefont {Jennings}, \citenamefont {Kanodia}, \citenamefont {Kaplan}, \citenamefont {Levi}, \citenamefont {Lubar}, \citenamefont {Mahadevan}, \citenamefont {Monson}, \citenamefont {Ninan}, \citenamefont {Nitroy}, \citenamefont {Osterman}, \citenamefont {Papp}, \citenamefont {Quinlan}, \citenamefont {Ramsey}, \citenamefont {Robertson}, \citenamefont {Roy}, \citenamefont {Schwab}, \citenamefont {Sigurdsson}, \citenamefont {Srinivasan}, \citenamefont {Stefansson}, \citenamefont {Sterner}, \citenamefont {Terrien}, \citenamefont {Wolszczan}, \citenamefont {Wright},\ and\ \citenamefont {Ycas}}]{MetcalfOpticaOPTICA2019a}%
  \BibitemOpen
  \bibfield  {author} {\bibinfo {author} {\bibfnamefont {A.~J.}\ \bibnamefont {Metcalf}}, \bibinfo {author} {\bibfnamefont {T.}~\bibnamefont {Anderson}}, \bibinfo {author} {\bibfnamefont {C.~F.}\ \bibnamefont {Bender}}, \bibinfo {author} {\bibfnamefont {S.}~\bibnamefont {Blakeslee}}, \bibinfo {author} {\bibfnamefont {W.}~\bibnamefont {Brand}}, \bibinfo {author} {\bibfnamefont {D.~R.}\ \bibnamefont {Carlson}}, \bibinfo {author} {\bibfnamefont {W.~D.}\ \bibnamefont {Cochran}}, \bibinfo {author} {\bibfnamefont {S.~A.}\ \bibnamefont {Diddams}}, \bibinfo {author} {\bibfnamefont {M.}~\bibnamefont {Endl}}, \bibinfo {author} {\bibfnamefont {C.}~\bibnamefont {Fredrick}}, \bibinfo {author} {\bibfnamefont {S.}~\bibnamefont {Halverson}}, \bibinfo {author} {\bibfnamefont {D.~D.}\ \bibnamefont {Hickstein}}, \bibinfo {author} {\bibfnamefont {F.}~\bibnamefont {Hearty}}, \bibinfo {author} {\bibfnamefont {J.}~\bibnamefont {Jennings}}, \bibinfo {author} {\bibfnamefont {S.}~\bibnamefont {Kanodia}}, \bibinfo {author} {\bibfnamefont {K.~F.}\ \bibnamefont {Kaplan}}, \bibinfo {author} {\bibfnamefont {E.}~\bibnamefont {Levi}}, \bibinfo {author} {\bibfnamefont {E.}~\bibnamefont {Lubar}}, \bibinfo {author} {\bibfnamefont {S.}~\bibnamefont {Mahadevan}}, \bibinfo {author} {\bibfnamefont {A.}~\bibnamefont {Monson}}, \bibinfo {author} {\bibfnamefont {J.~P.}\ \bibnamefont {Ninan}}, \bibinfo {author} {\bibfnamefont {C.}~\bibnamefont {Nitroy}}, \bibinfo {author} {\bibfnamefont {S.}~\bibnamefont {Osterman}}, \bibinfo {author} {\bibfnamefont {S.~B.}\ \bibnamefont {Papp}}, \bibinfo {author} {\bibfnamefont {F.}~\bibnamefont {Quinlan}}, \bibinfo {author} {\bibfnamefont {L.}~\bibnamefont {Ramsey}}, \bibinfo {author} {\bibfnamefont {P.}~\bibnamefont {Robertson}}, \bibinfo {author} {\bibfnamefont {A.}~\bibnamefont {Roy}}, \bibinfo {author} {\bibfnamefont {C.}~\bibnamefont {Schwab}}, \bibinfo {author} {\bibfnamefont {S.}~\bibnamefont {Sigurdsson}}, \bibinfo {author} {\bibfnamefont {K.}~\bibnamefont
  {Srinivasan}}, \bibinfo {author} {\bibfnamefont {G.}~\bibnamefont {Stefansson}}, \bibinfo {author} {\bibfnamefont {D.~A.}\ \bibnamefont {Sterner}}, \bibinfo {author} {\bibfnamefont {R.}~\bibnamefont {Terrien}}, \bibinfo {author} {\bibfnamefont {A.}~\bibnamefont {Wolszczan}}, \bibinfo {author} {\bibfnamefont {J.~T.}\ \bibnamefont {Wright}},\ and\ \bibinfo {author} {\bibfnamefont {G.}~\bibnamefont {Ycas}},\ }\bibfield  {title} {\enquote {\bibinfo {title} {Stellar spectroscopy in the near-infrared with a laser frequency comb},}\ }\href {https://doi.org/10.1364/OPTICA.6.000233} {\bibfield  {journal} {\bibinfo  {journal} {Optica}\ }\textbf {\bibinfo {volume} {6}},\ \bibinfo {pages} {233--239} (\bibinfo {year} {2019})}\BibitemShut {NoStop}%
\bibitem [{\citenamefont {Fortier}\ \emph {et~al.}(2011)\citenamefont {Fortier}, \citenamefont {Kirchner}, \citenamefont {Quinlan}, \citenamefont {Taylor}, \citenamefont {Bergquist}, \citenamefont {Rosenband}, \citenamefont {Lemke}, \citenamefont {Ludlow}, \citenamefont {Jiang}, \citenamefont {Oates},\ and\ \citenamefont {Diddams}}]{FortierNaturePhoton2011}%
  \BibitemOpen
  \bibfield  {author} {\bibinfo {author} {\bibfnamefont {T.~M.}\ \bibnamefont {Fortier}}, \bibinfo {author} {\bibfnamefont {M.~S.}\ \bibnamefont {Kirchner}}, \bibinfo {author} {\bibfnamefont {F.}~\bibnamefont {Quinlan}}, \bibinfo {author} {\bibfnamefont {J.}~\bibnamefont {Taylor}}, \bibinfo {author} {\bibfnamefont {J.~C.}\ \bibnamefont {Bergquist}}, \bibinfo {author} {\bibfnamefont {T.}~\bibnamefont {Rosenband}}, \bibinfo {author} {\bibfnamefont {N.}~\bibnamefont {Lemke}}, \bibinfo {author} {\bibfnamefont {A.}~\bibnamefont {Ludlow}}, \bibinfo {author} {\bibfnamefont {Y.}~\bibnamefont {Jiang}}, \bibinfo {author} {\bibfnamefont {C.~W.}\ \bibnamefont {Oates}},\ and\ \bibinfo {author} {\bibfnamefont {S.~A.}\ \bibnamefont {Diddams}},\ }\bibfield  {title} {\enquote {\bibinfo {title} {Generation of ultrastable microwaves via optical frequency division},}\ }\href {https://doi.org/10.1038/nphoton.2011.121} {\bibfield  {journal} {\bibinfo  {journal} {Nature Photonics}\ }\textbf {\bibinfo {volume} {5}},\ \bibinfo {pages} {425--429} (\bibinfo {year} {2011})}\BibitemShut {NoStop}%
\bibitem [{\citenamefont {Holzwarth}\ \emph {et~al.}(2000)\citenamefont {Holzwarth}, \citenamefont {Udem}, \citenamefont {H{\"a}nsch}, \citenamefont {Knight}, \citenamefont {Wadsworth},\ and\ \citenamefont {Russell}}]{HolzwarthPhys.Rev.Lett.2000}%
  \BibitemOpen
  \bibfield  {author} {\bibinfo {author} {\bibfnamefont {R.}~\bibnamefont {Holzwarth}}, \bibinfo {author} {\bibfnamefont {{\relax Th}.}~\bibnamefont {Udem}}, \bibinfo {author} {\bibfnamefont {T.~W.}\ \bibnamefont {H{\"a}nsch}}, \bibinfo {author} {\bibfnamefont {J.~C.}\ \bibnamefont {Knight}}, \bibinfo {author} {\bibfnamefont {W.~J.}\ \bibnamefont {Wadsworth}},\ and\ \bibinfo {author} {\bibfnamefont {P.~{\relax St}.~J.}\ \bibnamefont {Russell}},\ }\bibfield  {title} {\enquote {\bibinfo {title} {Optical {{Frequency Synthesizer}} for {{Precision Spectroscopy}}},}\ }\href {https://doi.org/10.1103/PhysRevLett.85.2264} {\bibfield  {journal} {\bibinfo  {journal} {Physical Review Letters}\ }\textbf {\bibinfo {volume} {85}},\ \bibinfo {pages} {2264--2267} (\bibinfo {year} {2000})}\BibitemShut {NoStop}%
\bibitem [{\citenamefont {Moody}\ \emph {et~al.}(2020)\citenamefont {Moody}, \citenamefont {Chang}, \citenamefont {Steiner},\ and\ \citenamefont {Bowers}}]{MoodyAVSQuantumSci.2020}%
  \BibitemOpen
  \bibfield  {author} {\bibinfo {author} {\bibfnamefont {G.}~\bibnamefont {Moody}}, \bibinfo {author} {\bibfnamefont {L.}~\bibnamefont {Chang}}, \bibinfo {author} {\bibfnamefont {T.~J.}\ \bibnamefont {Steiner}},\ and\ \bibinfo {author} {\bibfnamefont {J.~E.}\ \bibnamefont {Bowers}},\ }\bibfield  {title} {\enquote {\bibinfo {title} {Chip-scale nonlinear photonics for quantum light generation},}\ }\href {https://doi.org/10.1116/5.0020684} {\bibfield  {journal} {\bibinfo  {journal} {AVS Quantum Science}\ }\textbf {\bibinfo {volume} {2}},\ \bibinfo {pages} {041702} (\bibinfo {year} {2020})}\BibitemShut {NoStop}%
\bibitem [{\citenamefont {Stern}\ \emph {et~al.}(2018)\citenamefont {Stern}, \citenamefont {Ji}, \citenamefont {Okawachi}, \citenamefont {Gaeta},\ and\ \citenamefont {Lipson}}]{SternNature2018a}%
  \BibitemOpen
  \bibfield  {author} {\bibinfo {author} {\bibfnamefont {B.}~\bibnamefont {Stern}}, \bibinfo {author} {\bibfnamefont {X.}~\bibnamefont {Ji}}, \bibinfo {author} {\bibfnamefont {Y.}~\bibnamefont {Okawachi}}, \bibinfo {author} {\bibfnamefont {A.~L.}\ \bibnamefont {Gaeta}},\ and\ \bibinfo {author} {\bibfnamefont {M.}~\bibnamefont {Lipson}},\ }\bibfield  {title} {\enquote {\bibinfo {title} {Battery-operated integrated frequency comb generator},}\ }\href {https://doi.org/10.1038/s41586-018-0598-9} {\bibfield  {journal} {\bibinfo  {journal} {Nature}\ }\textbf {\bibinfo {volume} {562}},\ \bibinfo {pages} {401--405} (\bibinfo {year} {2018})}\BibitemShut {NoStop}%
\bibitem [{\citenamefont {Zhang}\ \emph {et~al.}(2019{\natexlab{a}})\citenamefont {Zhang}, \citenamefont {Buscaino}, \citenamefont {Wang}, \citenamefont {{Shams-Ansari}}, \citenamefont {Reimer}, \citenamefont {Zhu}, \citenamefont {Kahn},\ and\ \citenamefont {Lon{\v c}ar}}]{ZhangNature2019}%
  \BibitemOpen
  \bibfield  {author} {\bibinfo {author} {\bibfnamefont {M.}~\bibnamefont {Zhang}}, \bibinfo {author} {\bibfnamefont {B.}~\bibnamefont {Buscaino}}, \bibinfo {author} {\bibfnamefont {C.}~\bibnamefont {Wang}}, \bibinfo {author} {\bibfnamefont {A.}~\bibnamefont {{Shams-Ansari}}}, \bibinfo {author} {\bibfnamefont {C.}~\bibnamefont {Reimer}}, \bibinfo {author} {\bibfnamefont {R.}~\bibnamefont {Zhu}}, \bibinfo {author} {\bibfnamefont {J.~M.}\ \bibnamefont {Kahn}},\ and\ \bibinfo {author} {\bibfnamefont {M.}~\bibnamefont {Lon{\v c}ar}},\ }\bibfield  {title} {\enquote {\bibinfo {title} {Broadband electro-optic frequency comb generation in a lithium niobate microring resonator},}\ }\href {https://doi.org/10.1038/s41586-019-1008-7} {\bibfield  {journal} {\bibinfo  {journal} {Nature}\ }\textbf {\bibinfo {volume} {568}},\ \bibinfo {pages} {373--377} (\bibinfo {year} {2019}{\natexlab{a}})}\BibitemShut {NoStop}%
\bibitem [{\citenamefont {Wang}\ \emph {et~al.}(2024)\citenamefont {Wang}, \citenamefont {Li}, \citenamefont {Riemensberger}, \citenamefont {Lihachev}, \citenamefont {Churaev}, \citenamefont {Kao}, \citenamefont {Ji}, \citenamefont {Zhang}, \citenamefont {Blesin}, \citenamefont {Davydova}, \citenamefont {Chen}, \citenamefont {Huang}, \citenamefont {Wang}, \citenamefont {Ou},\ and\ \citenamefont {Kippenberg}}]{WangNature2024}%
  \BibitemOpen
  \bibfield  {author} {\bibinfo {author} {\bibfnamefont {C.}~\bibnamefont {Wang}}, \bibinfo {author} {\bibfnamefont {Z.}~\bibnamefont {Li}}, \bibinfo {author} {\bibfnamefont {J.}~\bibnamefont {Riemensberger}}, \bibinfo {author} {\bibfnamefont {G.}~\bibnamefont {Lihachev}}, \bibinfo {author} {\bibfnamefont {M.}~\bibnamefont {Churaev}}, \bibinfo {author} {\bibfnamefont {W.}~\bibnamefont {Kao}}, \bibinfo {author} {\bibfnamefont {X.}~\bibnamefont {Ji}}, \bibinfo {author} {\bibfnamefont {J.}~\bibnamefont {Zhang}}, \bibinfo {author} {\bibfnamefont {T.}~\bibnamefont {Blesin}}, \bibinfo {author} {\bibfnamefont {A.}~\bibnamefont {Davydova}}, \bibinfo {author} {\bibfnamefont {Y.}~\bibnamefont {Chen}}, \bibinfo {author} {\bibfnamefont {K.}~\bibnamefont {Huang}}, \bibinfo {author} {\bibfnamefont {X.}~\bibnamefont {Wang}}, \bibinfo {author} {\bibfnamefont {X.}~\bibnamefont {Ou}},\ and\ \bibinfo {author} {\bibfnamefont {T.~J.}\ \bibnamefont {Kippenberg}},\ }\bibfield  {title} {\enquote {\bibinfo {title} {Lithium tantalate photonic integrated circuits for volume manufacturing},}\ }\href {https://doi.org/10.1038/s41586-024-07369-1} {\bibfield  {journal} {\bibinfo  {journal} {Nature}\ }\textbf {\bibinfo {volume} {629}},\ \bibinfo {pages} {784--790} (\bibinfo {year} {2024})}\BibitemShut {NoStop}%
\bibitem [{\citenamefont {Liu}\ \emph {et~al.}(2021)\citenamefont {Liu}, \citenamefont {Huang}, \citenamefont {Wang}, \citenamefont {He}, \citenamefont {Raja}, \citenamefont {Liu}, \citenamefont {Engelsen},\ and\ \citenamefont {Kippenberg}}]{LiuNatCommun2021a}%
  \BibitemOpen
  \bibfield  {author} {\bibinfo {author} {\bibfnamefont {J.}~\bibnamefont {Liu}}, \bibinfo {author} {\bibfnamefont {G.}~\bibnamefont {Huang}}, \bibinfo {author} {\bibfnamefont {R.~N.}\ \bibnamefont {Wang}}, \bibinfo {author} {\bibfnamefont {J.}~\bibnamefont {He}}, \bibinfo {author} {\bibfnamefont {A.~S.}\ \bibnamefont {Raja}}, \bibinfo {author} {\bibfnamefont {T.}~\bibnamefont {Liu}}, \bibinfo {author} {\bibfnamefont {N.~J.}\ \bibnamefont {Engelsen}},\ and\ \bibinfo {author} {\bibfnamefont {T.~J.}\ \bibnamefont {Kippenberg}},\ }\bibfield  {title} {\enquote {\bibinfo {title} {High-yield, wafer-scale fabrication of ultralow-loss, dispersion-engineered silicon nitride photonic circuits},}\ }\href {https://doi.org/10.1038/s41467-021-21973-z} {\bibfield  {journal} {\bibinfo  {journal} {Nature Communications}\ }\textbf {\bibinfo {volume} {12}},\ \bibinfo {pages} {2236} (\bibinfo {year} {2021})}\BibitemShut {NoStop}%
\bibitem [{\citenamefont {Ou}\ \emph {et~al.}(2025)\citenamefont {Ou}, \citenamefont {Antohe}, \citenamefont {Carpenter}, \citenamefont {Moille},\ and\ \citenamefont {Srinivasan}}]{OuOpt.Lett.OL2025}%
  \BibitemOpen
  \bibfield  {author} {\bibinfo {author} {\bibfnamefont {S.-C.}\ \bibnamefont {Ou}}, \bibinfo {author} {\bibfnamefont {A.~O.}\ \bibnamefont {Antohe}}, \bibinfo {author} {\bibfnamefont {L.~G.}\ \bibnamefont {Carpenter}}, \bibinfo {author} {\bibfnamefont {G.}~\bibnamefont {Moille}},\ and\ \bibinfo {author} {\bibfnamefont {K.}~\bibnamefont {Srinivasan}},\ }\bibfield  {title} {\enquote {\bibinfo {title} {300 mm wafer-scale {{SiN}} platform for broadband soliton microcombs compatible with alkali atomic references},}\ }\href {https://doi.org/10.1364/OL.571893} {\bibfield  {journal} {\bibinfo  {journal} {Optics Letters}\ }\textbf {\bibinfo {volume} {50}},\ \bibinfo {pages} {5578--5581} (\bibinfo {year} {2025})}\BibitemShut {NoStop}%
\bibitem [{\citenamefont {Kippenberg}\ \emph {et~al.}(2018)\citenamefont {Kippenberg}, \citenamefont {Gaeta}, \citenamefont {Lipson},\ and\ \citenamefont {Gorodetsky}}]{KippenbergScience2018}%
  \BibitemOpen
  \bibfield  {author} {\bibinfo {author} {\bibfnamefont {T.~J.}\ \bibnamefont {Kippenberg}}, \bibinfo {author} {\bibfnamefont {A.~L.}\ \bibnamefont {Gaeta}}, \bibinfo {author} {\bibfnamefont {M.}~\bibnamefont {Lipson}},\ and\ \bibinfo {author} {\bibfnamefont {M.~L.}\ \bibnamefont {Gorodetsky}},\ }\bibfield  {title} {\enquote {\bibinfo {title} {Dissipative {{Kerr}} solitons in optical microresonators},}\ }\href {https://doi.org/10.1126/science.aan8083} {\bibfield  {journal} {\bibinfo  {journal} {Science}\ }\textbf {\bibinfo {volume} {361}},\ \bibinfo {pages} {eaan8083} (\bibinfo {year} {2018})}\BibitemShut {NoStop}%
\bibitem [{\citenamefont {Okawachi}\ \emph {et~al.}(2014)\citenamefont {Okawachi}, \citenamefont {Lamont}, \citenamefont {Luke}, \citenamefont {Carvalho}, \citenamefont {Yu}, \citenamefont {Lipson},\ and\ \citenamefont {Gaeta}}]{OkawachiOpt.Lett.2014}%
  \BibitemOpen
  \bibfield  {author} {\bibinfo {author} {\bibfnamefont {Y.}~\bibnamefont {Okawachi}}, \bibinfo {author} {\bibfnamefont {M.~R.~E.}\ \bibnamefont {Lamont}}, \bibinfo {author} {\bibfnamefont {K.}~\bibnamefont {Luke}}, \bibinfo {author} {\bibfnamefont {D.~O.}\ \bibnamefont {Carvalho}}, \bibinfo {author} {\bibfnamefont {M.}~\bibnamefont {Yu}}, \bibinfo {author} {\bibfnamefont {M.}~\bibnamefont {Lipson}},\ and\ \bibinfo {author} {\bibfnamefont {A.~L.}\ \bibnamefont {Gaeta}},\ }\bibfield  {title} {\enquote {\bibinfo {title} {Bandwidth shaping of microresonator-based frequency combs via dispersion engineering},}\ }\href {https://doi.org/10.1364/OL.39.003535} {\bibfield  {journal} {\bibinfo  {journal} {Optics Letters}\ }\textbf {\bibinfo {volume} {39}},\ \bibinfo {pages} {3535} (\bibinfo {year} {2014})}\BibitemShut {NoStop}%
\bibitem [{\citenamefont {Li}\ \emph {et~al.}(2017)\citenamefont {Li}, \citenamefont {Briles}, \citenamefont {Westly}, \citenamefont {Drake}, \citenamefont {Stone}, \citenamefont {Ilic}, \citenamefont {Diddams}, \citenamefont {Papp},\ and\ \citenamefont {Srinivasan}}]{LiOpticaOPTICA2017}%
  \BibitemOpen
  \bibfield  {author} {\bibinfo {author} {\bibfnamefont {Q.}~\bibnamefont {Li}}, \bibinfo {author} {\bibfnamefont {T.~C.}\ \bibnamefont {Briles}}, \bibinfo {author} {\bibfnamefont {D.~A.}\ \bibnamefont {Westly}}, \bibinfo {author} {\bibfnamefont {T.~E.}\ \bibnamefont {Drake}}, \bibinfo {author} {\bibfnamefont {J.~R.}\ \bibnamefont {Stone}}, \bibinfo {author} {\bibfnamefont {B.~R.}\ \bibnamefont {Ilic}}, \bibinfo {author} {\bibfnamefont {S.~A.}\ \bibnamefont {Diddams}}, \bibinfo {author} {\bibfnamefont {S.~B.}\ \bibnamefont {Papp}},\ and\ \bibinfo {author} {\bibfnamefont {K.}~\bibnamefont {Srinivasan}},\ }\bibfield  {title} {\enquote {\bibinfo {title} {Stably accessing octave-spanning microresonator frequency combs in the soliton regime},}\ }\href {https://doi.org/10.1364/OPTICA.4.000193} {\bibfield  {journal} {\bibinfo  {journal} {Optica}\ }\textbf {\bibinfo {volume} {4}},\ \bibinfo {pages} {193--203} (\bibinfo {year} {2017})}\BibitemShut {NoStop}%
\bibitem [{\citenamefont {Spencer}\ \emph {et~al.}(2018)\citenamefont {Spencer}, \citenamefont {Drake}, \citenamefont {Briles}, \citenamefont {Stone}, \citenamefont {Sinclair}, \citenamefont {Fredrick}, \citenamefont {Li}, \citenamefont {Westly}, \citenamefont {Ilic}, \citenamefont {Bluestone}, \citenamefont {Volet}, \citenamefont {Komljenovic}, \citenamefont {Chang}, \citenamefont {Lee}, \citenamefont {Oh}, \citenamefont {Suh}, \citenamefont {Yang}, \citenamefont {Pfeiffer}, \citenamefont {Kippenberg}, \citenamefont {Norberg}, \citenamefont {Theogarajan}, \citenamefont {Vahala}, \citenamefont {Newbury}, \citenamefont {Srinivasan}, \citenamefont {Bowers}, \citenamefont {Diddams},\ and\ \citenamefont {Papp}}]{SpencerNature2018}%
  \BibitemOpen
  \bibfield  {author} {\bibinfo {author} {\bibfnamefont {D.~T.}\ \bibnamefont {Spencer}}, \bibinfo {author} {\bibfnamefont {T.}~\bibnamefont {Drake}}, \bibinfo {author} {\bibfnamefont {T.~C.}\ \bibnamefont {Briles}}, \bibinfo {author} {\bibfnamefont {J.}~\bibnamefont {Stone}}, \bibinfo {author} {\bibfnamefont {L.~C.}\ \bibnamefont {Sinclair}}, \bibinfo {author} {\bibfnamefont {C.}~\bibnamefont {Fredrick}}, \bibinfo {author} {\bibfnamefont {Q.}~\bibnamefont {Li}}, \bibinfo {author} {\bibfnamefont {D.}~\bibnamefont {Westly}}, \bibinfo {author} {\bibfnamefont {B.~R.}\ \bibnamefont {Ilic}}, \bibinfo {author} {\bibfnamefont {A.}~\bibnamefont {Bluestone}}, \bibinfo {author} {\bibfnamefont {N.}~\bibnamefont {Volet}}, \bibinfo {author} {\bibfnamefont {T.}~\bibnamefont {Komljenovic}}, \bibinfo {author} {\bibfnamefont {L.}~\bibnamefont {Chang}}, \bibinfo {author} {\bibfnamefont {S.~H.}\ \bibnamefont {Lee}}, \bibinfo {author} {\bibfnamefont {D.~Y.}\ \bibnamefont {Oh}}, \bibinfo {author} {\bibfnamefont {M.-G.}\ \bibnamefont {Suh}}, \bibinfo {author} {\bibfnamefont {K.~Y.}\ \bibnamefont {Yang}}, \bibinfo {author} {\bibfnamefont {M.~H.~P.}\ \bibnamefont {Pfeiffer}}, \bibinfo {author} {\bibfnamefont {T.~J.}\ \bibnamefont {Kippenberg}}, \bibinfo {author} {\bibfnamefont {E.}~\bibnamefont {Norberg}}, \bibinfo {author} {\bibfnamefont {L.}~\bibnamefont {Theogarajan}}, \bibinfo {author} {\bibfnamefont {K.}~\bibnamefont {Vahala}}, \bibinfo {author} {\bibfnamefont {N.~R.}\ \bibnamefont {Newbury}}, \bibinfo {author} {\bibfnamefont {K.}~\bibnamefont {Srinivasan}}, \bibinfo {author} {\bibfnamefont {J.~E.}\ \bibnamefont {Bowers}}, \bibinfo {author} {\bibfnamefont {S.~A.}\ \bibnamefont {Diddams}},\ and\ \bibinfo {author} {\bibfnamefont {S.~B.}\ \bibnamefont {Papp}},\ }\bibfield  {title} {\enquote {\bibinfo {title} {An optical-frequency synthesizer using integrated photonics},}\ }\href {https://doi.org/10.1038/s41586-018-0065-7} {\bibfield  {journal} {\bibinfo  {journal} {Nature}\ }\textbf
  {\bibinfo {volume} {557}},\ \bibinfo {pages} {81--85} (\bibinfo {year} {2018})}\BibitemShut {NoStop}%
\bibitem [{\citenamefont {Newman}\ \emph {et~al.}(2019)\citenamefont {Newman}, \citenamefont {Maurice}, \citenamefont {Drake}, \citenamefont {Stone}, \citenamefont {Briles}, \citenamefont {Spencer}, \citenamefont {Fredrick}, \citenamefont {Li}, \citenamefont {Westly}, \citenamefont {Ilic}, \citenamefont {Shen}, \citenamefont {Suh}, \citenamefont {Yang}, \citenamefont {Johnson}, \citenamefont {Johnson}, \citenamefont {Hollberg}, \citenamefont {Vahala}, \citenamefont {Srinivasan}, \citenamefont {Diddams}, \citenamefont {Kitching}, \citenamefont {Papp},\ and\ \citenamefont {Hummon}}]{NewmanOptica2019}%
  \BibitemOpen
  \bibfield  {author} {\bibinfo {author} {\bibfnamefont {Z.~L.}\ \bibnamefont {Newman}}, \bibinfo {author} {\bibfnamefont {V.}~\bibnamefont {Maurice}}, \bibinfo {author} {\bibfnamefont {T.}~\bibnamefont {Drake}}, \bibinfo {author} {\bibfnamefont {J.~R.}\ \bibnamefont {Stone}}, \bibinfo {author} {\bibfnamefont {T.~C.}\ \bibnamefont {Briles}}, \bibinfo {author} {\bibfnamefont {D.~T.}\ \bibnamefont {Spencer}}, \bibinfo {author} {\bibfnamefont {C.}~\bibnamefont {Fredrick}}, \bibinfo {author} {\bibfnamefont {Q.}~\bibnamefont {Li}}, \bibinfo {author} {\bibfnamefont {D.}~\bibnamefont {Westly}}, \bibinfo {author} {\bibfnamefont {B.~R.}\ \bibnamefont {Ilic}}, \bibinfo {author} {\bibfnamefont {B.}~\bibnamefont {Shen}}, \bibinfo {author} {\bibfnamefont {M.-G.}\ \bibnamefont {Suh}}, \bibinfo {author} {\bibfnamefont {K.~Y.}\ \bibnamefont {Yang}}, \bibinfo {author} {\bibfnamefont {C.}~\bibnamefont {Johnson}}, \bibinfo {author} {\bibfnamefont {D.~M.~S.}\ \bibnamefont {Johnson}}, \bibinfo {author} {\bibfnamefont {L.}~\bibnamefont {Hollberg}}, \bibinfo {author} {\bibfnamefont {K.~J.}\ \bibnamefont {Vahala}}, \bibinfo {author} {\bibfnamefont {K.}~\bibnamefont {Srinivasan}}, \bibinfo {author} {\bibfnamefont {S.~A.}\ \bibnamefont {Diddams}}, \bibinfo {author} {\bibfnamefont {J.}~\bibnamefont {Kitching}}, \bibinfo {author} {\bibfnamefont {S.~B.}\ \bibnamefont {Papp}},\ and\ \bibinfo {author} {\bibfnamefont {M.~T.}\ \bibnamefont {Hummon}},\ }\bibfield  {title} {\enquote {\bibinfo {title} {Architecture for the photonic integration of an optical atomic clock},}\ }\href {https://doi.org/10.1364/OPTICA.6.000680} {\bibfield  {journal} {\bibinfo  {journal} {Optica}\ }\textbf {\bibinfo {volume} {6}},\ \bibinfo {pages} {680} (\bibinfo {year} {2019})}\BibitemShut {NoStop}%
\bibitem [{\citenamefont {Moille}\ \emph {et~al.}(2023{\natexlab{a}})\citenamefont {Moille}, \citenamefont {Stone}, \citenamefont {Chojnacky}, \citenamefont {Shrestha}, \citenamefont {Javid}, \citenamefont {Menyuk},\ and\ \citenamefont {Srinivasan}}]{MoilleNature2023}%
  \BibitemOpen
  \bibfield  {author} {\bibinfo {author} {\bibfnamefont {G.}~\bibnamefont {Moille}}, \bibinfo {author} {\bibfnamefont {J.}~\bibnamefont {Stone}}, \bibinfo {author} {\bibfnamefont {M.}~\bibnamefont {Chojnacky}}, \bibinfo {author} {\bibfnamefont {R.}~\bibnamefont {Shrestha}}, \bibinfo {author} {\bibfnamefont {U.~A.}\ \bibnamefont {Javid}}, \bibinfo {author} {\bibfnamefont {C.}~\bibnamefont {Menyuk}},\ and\ \bibinfo {author} {\bibfnamefont {K.}~\bibnamefont {Srinivasan}},\ }\bibfield  {title} {\enquote {\bibinfo {title} {Kerr-induced synchronization of a cavity soliton to an optical reference},}\ }\href {https://doi.org/10.1038/s41586-023-06730-0} {\bibfield  {journal} {\bibinfo  {journal} {Nature}\ }\textbf {\bibinfo {volume} {624}},\ \bibinfo {pages} {267--274} (\bibinfo {year} {2023}{\natexlab{a}})}\BibitemShut {NoStop}%
\bibitem [{\citenamefont {Moille}\ \emph {et~al.}(2025{\natexlab{a}})\citenamefont {Moille}, \citenamefont {Shandilya}, \citenamefont {Stone}, \citenamefont {Menyuk},\ and\ \citenamefont {Srinivasan}}]{MoilleOptica2025}%
  \BibitemOpen
  \bibfield  {author} {\bibinfo {author} {\bibfnamefont {G.}~\bibnamefont {Moille}}, \bibinfo {author} {\bibfnamefont {P.}~\bibnamefont {Shandilya}}, \bibinfo {author} {\bibfnamefont {J.}~\bibnamefont {Stone}}, \bibinfo {author} {\bibfnamefont {C.}~\bibnamefont {Menyuk}},\ and\ \bibinfo {author} {\bibfnamefont {K.}~\bibnamefont {Srinivasan}},\ }\bibfield  {title} {\enquote {\bibinfo {title} {All-optical noise quenching of an integrated frequency comb},}\ }\href {https://doi.org/10.1364/optica.561954} {\bibfield  {journal} {\bibinfo  {journal} {Optica}\ }\textbf {\bibinfo {volume} {12}},\ \bibinfo {pages} {1020} (\bibinfo {year} {2025}{\natexlab{a}})}\BibitemShut {NoStop}%
\bibitem [{\citenamefont {Sun}\ \emph {et~al.}(2025)\citenamefont {Sun}, \citenamefont {Harrington}, \citenamefont {Tabatabaei}, \citenamefont {Hanifi}, \citenamefont {Liu}, \citenamefont {Wang}, \citenamefont {Wang}, \citenamefont {Yang}, \citenamefont {Liu}, \citenamefont {Morgan}, \citenamefont {Bowers}, \citenamefont {Morton}, \citenamefont {Nelson}, \citenamefont {Beling}, \citenamefont {Blumenthal},\ and\ \citenamefont {Yi}}]{SunNat.Photon.2025}%
  \BibitemOpen
  \bibfield  {author} {\bibinfo {author} {\bibfnamefont {S.}~\bibnamefont {Sun}}, \bibinfo {author} {\bibfnamefont {M.~W.}\ \bibnamefont {Harrington}}, \bibinfo {author} {\bibfnamefont {F.}~\bibnamefont {Tabatabaei}}, \bibinfo {author} {\bibfnamefont {S.}~\bibnamefont {Hanifi}}, \bibinfo {author} {\bibfnamefont {K.}~\bibnamefont {Liu}}, \bibinfo {author} {\bibfnamefont {J.}~\bibnamefont {Wang}}, \bibinfo {author} {\bibfnamefont {B.}~\bibnamefont {Wang}}, \bibinfo {author} {\bibfnamefont {Z.}~\bibnamefont {Yang}}, \bibinfo {author} {\bibfnamefont {R.}~\bibnamefont {Liu}}, \bibinfo {author} {\bibfnamefont {J.~S.}\ \bibnamefont {Morgan}}, \bibinfo {author} {\bibfnamefont {S.~M.}\ \bibnamefont {Bowers}}, \bibinfo {author} {\bibfnamefont {P.~A.}\ \bibnamefont {Morton}}, \bibinfo {author} {\bibfnamefont {K.~D.}\ \bibnamefont {Nelson}}, \bibinfo {author} {\bibfnamefont {A.}~\bibnamefont {Beling}}, \bibinfo {author} {\bibfnamefont {D.~J.}\ \bibnamefont {Blumenthal}},\ and\ \bibinfo {author} {\bibfnamefont {X.}~\bibnamefont {Yi}},\ }\bibfield  {title} {\enquote {\bibinfo {title} {Microcavity {{Kerr}} optical frequency division with integrated {{SiN}} photonics},}\ }\href {https://doi.org/10.1038/s41566-025-01668-3} {\bibfield  {journal} {\bibinfo  {journal} {Nature Photonics}\ ,\ \bibinfo {pages} {1--6}} (\bibinfo {year} {2025})}\BibitemShut {NoStop}%
\bibitem [{\citenamefont {Diakonov}, \citenamefont {Khrizman},\ and\ \citenamefont {Stern}(2025)}]{Diakonov2025}%
  \BibitemOpen
  \bibfield  {author} {\bibinfo {author} {\bibfnamefont {A.}~\bibnamefont {Diakonov}}, \bibinfo {author} {\bibfnamefont {K.}~\bibnamefont {Khrizman}},\ and\ \bibinfo {author} {\bibfnamefont {L.}~\bibnamefont {Stern}},\ }\href {https://doi.org/10.48550/arXiv.2508.07258} {\enquote {\bibinfo {title} {Hybrid-{{Locked Kerr Microcombs}} for {{Flexible On-Chip Optical Clock Division}}},}\ } (\bibinfo {year} {2025}),\ \Eprint {https://arxiv.org/abs/2508.07258} {arXiv:2508.07258 [physics]} \BibitemShut {NoStop}%
\bibitem [{\citenamefont {Menyuk}\ \emph {et~al.}(2025)\citenamefont {Menyuk}, \citenamefont {Shandilya}, \citenamefont {Courtright}, \citenamefont {Moille},\ and\ \citenamefont {Srinivasan}}]{MenyukOpt.ExpressOE2025}%
  \BibitemOpen
  \bibfield  {author} {\bibinfo {author} {\bibfnamefont {C.~R.}\ \bibnamefont {Menyuk}}, \bibinfo {author} {\bibfnamefont {P.}~\bibnamefont {Shandilya}}, \bibinfo {author} {\bibfnamefont {L.}~\bibnamefont {Courtright}}, \bibinfo {author} {\bibfnamefont {G.}~\bibnamefont {Moille}},\ and\ \bibinfo {author} {\bibfnamefont {K.}~\bibnamefont {Srinivasan}},\ }\bibfield  {title} {\enquote {\bibinfo {title} {Multi-color solitons and frequency combs in microresonators},}\ }\href {https://doi.org/10.1364/OE.544077} {\bibfield  {journal} {\bibinfo  {journal} {Optics Express}\ }\textbf {\bibinfo {volume} {33}},\ \bibinfo {pages} {21824--21835} (\bibinfo {year} {2025})}\BibitemShut {NoStop}%
\bibitem [{\citenamefont {Zhang}\ \emph {et~al.}(2020)\citenamefont {Zhang}, \citenamefont {Silver}, \citenamefont {Bi},\ and\ \citenamefont {Del'Haye}}]{ZhangNatCommun2020}%
  \BibitemOpen
  \bibfield  {author} {\bibinfo {author} {\bibfnamefont {S.}~\bibnamefont {Zhang}}, \bibinfo {author} {\bibfnamefont {J.~M.}\ \bibnamefont {Silver}}, \bibinfo {author} {\bibfnamefont {T.}~\bibnamefont {Bi}},\ and\ \bibinfo {author} {\bibfnamefont {P.}~\bibnamefont {Del'Haye}},\ }\bibfield  {title} {\enquote {\bibinfo {title} {Spectral extension and synchronization of microcombs in a single microresonator},}\ }\href {https://doi.org/10.1038/s41467-020-19804-8} {\bibfield  {journal} {\bibinfo  {journal} {Nature Communications}\ }\textbf {\bibinfo {volume} {11}},\ \bibinfo {pages} {6384} (\bibinfo {year} {2020})}\BibitemShut {NoStop}%
\bibitem [{\citenamefont {Moille}\ \emph {et~al.}(2021)\citenamefont {Moille}, \citenamefont {Perez}, \citenamefont {Stone}, \citenamefont {Rao}, \citenamefont {Lu}, \citenamefont {Rahman}, \citenamefont {Chembo},\ and\ \citenamefont {Srinivasan}}]{MoilleNat.Commun.2021}%
  \BibitemOpen
  \bibfield  {author} {\bibinfo {author} {\bibfnamefont {G.}~\bibnamefont {Moille}}, \bibinfo {author} {\bibfnamefont {E.~F.}\ \bibnamefont {Perez}}, \bibinfo {author} {\bibfnamefont {J.~R.}\ \bibnamefont {Stone}}, \bibinfo {author} {\bibfnamefont {A.}~\bibnamefont {Rao}}, \bibinfo {author} {\bibfnamefont {X.}~\bibnamefont {Lu}}, \bibinfo {author} {\bibfnamefont {T.~S.}\ \bibnamefont {Rahman}}, \bibinfo {author} {\bibfnamefont {Y.~K.}\ \bibnamefont {Chembo}},\ and\ \bibinfo {author} {\bibfnamefont {K.}~\bibnamefont {Srinivasan}},\ }\bibfield  {title} {\enquote {\bibinfo {title} {Ultra-broadband {{Kerr}} microcomb through soliton spectral translation},}\ }\href {https://doi.org/10.1038/s41467-021-27469-0} {\bibfield  {journal} {\bibinfo  {journal} {Nature Communications}\ }\textbf {\bibinfo {volume} {12}},\ \bibinfo {pages} {7275} (\bibinfo {year} {2021})}\BibitemShut {NoStop}%
\bibitem [{\citenamefont {Zhang}\ \emph {et~al.}(2022)\citenamefont {Zhang}, \citenamefont {Bi}, \citenamefont {Ghalanos}, \citenamefont {Moroney}, \citenamefont {Del~Bino},\ and\ \citenamefont {Del'Haye}}]{ZhangPhys.Rev.Lett.2022}%
  \BibitemOpen
  \bibfield  {author} {\bibinfo {author} {\bibfnamefont {S.}~\bibnamefont {Zhang}}, \bibinfo {author} {\bibfnamefont {T.}~\bibnamefont {Bi}}, \bibinfo {author} {\bibfnamefont {G.~N.}\ \bibnamefont {Ghalanos}}, \bibinfo {author} {\bibfnamefont {N.~P.}\ \bibnamefont {Moroney}}, \bibinfo {author} {\bibfnamefont {L.}~\bibnamefont {Del~Bino}},\ and\ \bibinfo {author} {\bibfnamefont {P.}~\bibnamefont {Del'Haye}},\ }\bibfield  {title} {\enquote {\bibinfo {title} {Dark-{{Bright Soliton Bound States}} in a {{Microresonator}}},}\ }\href {https://doi.org/10.1103/PhysRevLett.128.033901} {\bibfield  {journal} {\bibinfo  {journal} {Physical Review Letters}\ }\textbf {\bibinfo {volume} {128}},\ \bibinfo {pages} {033901} (\bibinfo {year} {2022})}\BibitemShut {NoStop}%
\bibitem [{\citenamefont {Weng}\ \emph {et~al.}(2020)\citenamefont {Weng}, \citenamefont {Bouchand}, \citenamefont {Lucas}, \citenamefont {Obrzud}, \citenamefont {Herr},\ and\ \citenamefont {Kippenberg}}]{WengNatCommun2020a}%
  \BibitemOpen
  \bibfield  {author} {\bibinfo {author} {\bibfnamefont {W.}~\bibnamefont {Weng}}, \bibinfo {author} {\bibfnamefont {R.}~\bibnamefont {Bouchand}}, \bibinfo {author} {\bibfnamefont {E.}~\bibnamefont {Lucas}}, \bibinfo {author} {\bibfnamefont {E.}~\bibnamefont {Obrzud}}, \bibinfo {author} {\bibfnamefont {T.}~\bibnamefont {Herr}},\ and\ \bibinfo {author} {\bibfnamefont {T.~J.}\ \bibnamefont {Kippenberg}},\ }\bibfield  {title} {\enquote {\bibinfo {title} {Heteronuclear soliton molecules in optical microresonators},}\ }\href {https://doi.org/10.1038/s41467-020-15720-z} {\bibfield  {journal} {\bibinfo  {journal} {Nature Communications}\ }\textbf {\bibinfo {volume} {11}},\ \bibinfo {pages} {2402} (\bibinfo {year} {2020})}\BibitemShut {NoStop}%
\bibitem [{\citenamefont {Lucas}\ \emph {et~al.}(2025)\citenamefont {Lucas}, \citenamefont {Xu}, \citenamefont {Wang}, \citenamefont {Oppo}, \citenamefont {Hill}, \citenamefont {Del'Haye}, \citenamefont {Kibler}, \citenamefont {Xu}, \citenamefont {Murdoch}, \citenamefont {Erkintalo}, \citenamefont {Coen},\ and\ \citenamefont {Fatome}}]{LucasPhys.Rev.Lett.2025}%
  \BibitemOpen
  \bibfield  {author} {\bibinfo {author} {\bibfnamefont {E.}~\bibnamefont {Lucas}}, \bibinfo {author} {\bibfnamefont {G.}~\bibnamefont {Xu}}, \bibinfo {author} {\bibfnamefont {P.}~\bibnamefont {Wang}}, \bibinfo {author} {\bibfnamefont {G.-L.}\ \bibnamefont {Oppo}}, \bibinfo {author} {\bibfnamefont {L.}~\bibnamefont {Hill}}, \bibinfo {author} {\bibfnamefont {P.}~\bibnamefont {Del'Haye}}, \bibinfo {author} {\bibfnamefont {B.}~\bibnamefont {Kibler}}, \bibinfo {author} {\bibfnamefont {Y.}~\bibnamefont {Xu}}, \bibinfo {author} {\bibfnamefont {S.~G.}\ \bibnamefont {Murdoch}}, \bibinfo {author} {\bibfnamefont {M.}~\bibnamefont {Erkintalo}}, \bibinfo {author} {\bibfnamefont {S.}~\bibnamefont {Coen}},\ and\ \bibinfo {author} {\bibfnamefont {J.}~\bibnamefont {Fatome}},\ }\bibfield  {title} {\enquote {\bibinfo {title} {Polarization {{Faticons}}: {{Chiral Localized Structures}} in {{Self-Defocusing Kerr Resonators}}},}\ }\href {https://doi.org/10.1103/ljbj-tz7g} {\bibfield  {journal} {\bibinfo  {journal} {Physical Review Letters}\ }\textbf {\bibinfo {volume} {135}},\ \bibinfo {pages} {063803} (\bibinfo {year} {2025})}\BibitemShut {NoStop}%
\bibitem [{\citenamefont {Taheri}, \citenamefont {Matsko},\ and\ \citenamefont {Maleki}(2017)}]{TaheriEur.Phys.J.D2017}%
  \BibitemOpen
  \bibfield  {author} {\bibinfo {author} {\bibfnamefont {H.}~\bibnamefont {Taheri}}, \bibinfo {author} {\bibfnamefont {A.~B.}\ \bibnamefont {Matsko}},\ and\ \bibinfo {author} {\bibfnamefont {L.}~\bibnamefont {Maleki}},\ }\bibfield  {title} {\enquote {\bibinfo {title} {Optical lattice trap for {{Kerr}} solitons},}\ }\href {https://doi.org/10.1140/epjd/e2017-80150-6} {\bibfield  {journal} {\bibinfo  {journal} {The European Physical Journal D}\ }\textbf {\bibinfo {volume} {71}},\ \bibinfo {pages} {153} (\bibinfo {year} {2017})}\BibitemShut {NoStop}%
\bibitem [{\citenamefont {Li}, \citenamefont {Davan{\c c}o},\ and\ \citenamefont {Srinivasan}(2016)}]{LiNaturePhoton2016}%
  \BibitemOpen
  \bibfield  {author} {\bibinfo {author} {\bibfnamefont {Q.}~\bibnamefont {Li}}, \bibinfo {author} {\bibfnamefont {M.}~\bibnamefont {Davan{\c c}o}},\ and\ \bibinfo {author} {\bibfnamefont {K.}~\bibnamefont {Srinivasan}},\ }\bibfield  {title} {\enquote {\bibinfo {title} {Efficient and low-noise single-photon-level frequency conversion interfaces using silicon nanophotonics},}\ }\href {https://doi.org/10.1038/nphoton.2016.64} {\bibfield  {journal} {\bibinfo  {journal} {Nature Photonics}\ }\textbf {\bibinfo {volume} {10}},\ \bibinfo {pages} {406--414} (\bibinfo {year} {2016})}\BibitemShut {NoStop}%
\bibitem [{\citenamefont {Qureshi}\ \emph {et~al.}(2022)\citenamefont {Qureshi}, \citenamefont {Ng}, \citenamefont {Azeem}, \citenamefont {Trainor}, \citenamefont {Schwefel}, \citenamefont {Coen}, \citenamefont {Erkintalo},\ and\ \citenamefont {Murdoch}}]{QureshiCommunPhys2022}%
  \BibitemOpen
  \bibfield  {author} {\bibinfo {author} {\bibfnamefont {P.~C.}\ \bibnamefont {Qureshi}}, \bibinfo {author} {\bibfnamefont {V.}~\bibnamefont {Ng}}, \bibinfo {author} {\bibfnamefont {F.}~\bibnamefont {Azeem}}, \bibinfo {author} {\bibfnamefont {L.~S.}\ \bibnamefont {Trainor}}, \bibinfo {author} {\bibfnamefont {H.~G.~L.}\ \bibnamefont {Schwefel}}, \bibinfo {author} {\bibfnamefont {S.}~\bibnamefont {Coen}}, \bibinfo {author} {\bibfnamefont {M.}~\bibnamefont {Erkintalo}},\ and\ \bibinfo {author} {\bibfnamefont {S.~G.}\ \bibnamefont {Murdoch}},\ }\bibfield  {title} {\enquote {\bibinfo {title} {Soliton linear-wave scattering in a {{Kerr}} microresonator},}\ }\href {https://doi.org/10.1038/s42005-022-00903-5} {\bibfield  {journal} {\bibinfo  {journal} {Communications Physics}\ }\textbf {\bibinfo {volume} {5}},\ \bibinfo {pages} {1--8} (\bibinfo {year} {2022})}\BibitemShut {NoStop}%
\bibitem [{\citenamefont {Weng}\ \emph {et~al.}(2025)\citenamefont {Weng}, \citenamefont {Ji}, \citenamefont {Ali}, \citenamefont {Krock}, \citenamefont {Wang}, \citenamefont {Kumar}, \citenamefont {Guo}, \citenamefont {Kippenberg}, \citenamefont {Donegan},\ and\ \citenamefont {Skryabin}}]{Weng2025}%
  \BibitemOpen
  \bibfield  {author} {\bibinfo {author} {\bibfnamefont {H.}~\bibnamefont {Weng}}, \bibinfo {author} {\bibfnamefont {X.}~\bibnamefont {Ji}}, \bibinfo {author} {\bibfnamefont {M.}~\bibnamefont {Ali}}, \bibinfo {author} {\bibfnamefont {E.~H.}\ \bibnamefont {Krock}}, \bibinfo {author} {\bibfnamefont {L.}~\bibnamefont {Wang}}, \bibinfo {author} {\bibfnamefont {V.}~\bibnamefont {Kumar}}, \bibinfo {author} {\bibfnamefont {W.}~\bibnamefont {Guo}}, \bibinfo {author} {\bibfnamefont {T.~J.}\ \bibnamefont {Kippenberg}}, \bibinfo {author} {\bibfnamefont {J.~F.}\ \bibnamefont {Donegan}},\ and\ \bibinfo {author} {\bibfnamefont {D.~V.}\ \bibnamefont {Skryabin}},\ }\href {https://doi.org/10.48550/arXiv.2507.03626} {\enquote {\bibinfo {title} {Hyperparametric solitons in nondegenerate optical parametric oscillators},}\ } (\bibinfo {year} {2025}),\ \Eprint {https://arxiv.org/abs/2507.03626} {arXiv:2507.03626 [physics]} \BibitemShut {NoStop}%
\bibitem [{\citenamefont {Moille}\ \emph {et~al.}(2019)\citenamefont {Moille}, \citenamefont {Li}, \citenamefont {Xiyuan},\ and\ \citenamefont {Srinivasan}}]{MoilleJ.RES.NATL.INST.STAN.2019}%
  \BibitemOpen
  \bibfield  {author} {\bibinfo {author} {\bibfnamefont {G.}~\bibnamefont {Moille}}, \bibinfo {author} {\bibfnamefont {Q.}~\bibnamefont {Li}}, \bibinfo {author} {\bibfnamefont {L.}~\bibnamefont {Xiyuan}},\ and\ \bibinfo {author} {\bibfnamefont {K.}~\bibnamefont {Srinivasan}},\ }\bibfield  {title} {\enquote {\bibinfo {title} {{{pyLLE}}: {{A Fast}} and {{User Friendly Lugiato-Lefever Equation Solver}}},}\ }\href {https://doi.org/10.6028/jres.124.012} {\bibfield  {journal} {\bibinfo  {journal} {Journal of Research of the NIST}\ }\textbf {\bibinfo {volume} {124}},\ \bibinfo {pages} {124012} (\bibinfo {year} {2019})}\BibitemShut {NoStop}%
\bibitem [{\citenamefont {Miri}\ and\ \citenamefont {Al{\`u}}(2016)}]{MiriNewJ.Phys.2016}%
  \BibitemOpen
  \bibfield  {author} {\bibinfo {author} {\bibfnamefont {M.-A.}\ \bibnamefont {Miri}}\ and\ \bibinfo {author} {\bibfnamefont {A.}~\bibnamefont {Al{\`u}}},\ }\bibfield  {title} {\enquote {\bibinfo {title} {Nonlinearity-induced {{PT-symmetry}} without material gain},}\ }\href {https://doi.org/10.1088/1367-2630/18/6/065001} {\bibfield  {journal} {\bibinfo  {journal} {New Journal of Physics}\ }\textbf {\bibinfo {volume} {18}},\ \bibinfo {pages} {065001} (\bibinfo {year} {2016})}\BibitemShut {NoStop}%
\bibitem [{\citenamefont {Park}\ \emph {et~al.}(2021)\citenamefont {Park}, \citenamefont {Lee}, \citenamefont {Park}, \citenamefont {Shin}, \citenamefont {Choi},\ and\ \citenamefont {Yoon}}]{ParkPhys.Rev.Lett.2021}%
  \BibitemOpen
  \bibfield  {author} {\bibinfo {author} {\bibfnamefont {S.}~\bibnamefont {Park}}, \bibinfo {author} {\bibfnamefont {D.}~\bibnamefont {Lee}}, \bibinfo {author} {\bibfnamefont {K.}~\bibnamefont {Park}}, \bibinfo {author} {\bibfnamefont {H.}~\bibnamefont {Shin}}, \bibinfo {author} {\bibfnamefont {Y.}~\bibnamefont {Choi}},\ and\ \bibinfo {author} {\bibfnamefont {J.~W.}\ \bibnamefont {Yoon}},\ }\bibfield  {title} {\enquote {\bibinfo {title} {Optical {{Energy-Difference Conservation}} in a {{Synthetic Anti-PT-Symmetric System}}},}\ }\href {https://doi.org/10.1103/PhysRevLett.127.083601} {\bibfield  {journal} {\bibinfo  {journal} {Physical Review Letters}\ }\textbf {\bibinfo {volume} {127}},\ \bibinfo {pages} {083601} (\bibinfo {year} {2021})}\BibitemShut {NoStop}%
\bibitem [{\citenamefont {Zhang}\ \emph {et~al.}(2019{\natexlab{b}})\citenamefont {Zhang}, \citenamefont {Silver}, \citenamefont {Del~Bino}, \citenamefont {Copie}, \citenamefont {Woodley}, \citenamefont {Ghalanos}, \citenamefont {Svela}, \citenamefont {Moroney},\ and\ \citenamefont {Del'Haye}}]{ZhangOptica2019}%
  \BibitemOpen
  \bibfield  {author} {\bibinfo {author} {\bibfnamefont {S.}~\bibnamefont {Zhang}}, \bibinfo {author} {\bibfnamefont {J.~M.}\ \bibnamefont {Silver}}, \bibinfo {author} {\bibfnamefont {L.}~\bibnamefont {Del~Bino}}, \bibinfo {author} {\bibfnamefont {F.}~\bibnamefont {Copie}}, \bibinfo {author} {\bibfnamefont {M.~T.~M.}\ \bibnamefont {Woodley}}, \bibinfo {author} {\bibfnamefont {G.~N.}\ \bibnamefont {Ghalanos}}, \bibinfo {author} {\bibfnamefont {A.~{\O}.}\ \bibnamefont {Svela}}, \bibinfo {author} {\bibfnamefont {N.}~\bibnamefont {Moroney}},\ and\ \bibinfo {author} {\bibfnamefont {P.}~\bibnamefont {Del'Haye}},\ }\bibfield  {title} {\enquote {\bibinfo {title} {Sub-milliwatt-level microresonator solitons with extended access range using an auxiliary laser},}\ }\href {https://doi.org/10.1364/OPTICA.6.000206} {\bibfield  {journal} {\bibinfo  {journal} {Optica}\ }\textbf {\bibinfo {volume} {6}},\ \bibinfo {pages} {206} (\bibinfo {year} {2019}{\natexlab{b}})}\BibitemShut {NoStop}%
\bibitem [{\citenamefont {Zhou}\ \emph {et~al.}(2019)\citenamefont {Zhou}, \citenamefont {Geng}, \citenamefont {Cui}, \citenamefont {Huang}, \citenamefont {Zhou}, \citenamefont {Qiu},\ and\ \citenamefont {Wei~Wong}}]{ZhouLightSciAppl2019}%
  \BibitemOpen
  \bibfield  {author} {\bibinfo {author} {\bibfnamefont {H.}~\bibnamefont {Zhou}}, \bibinfo {author} {\bibfnamefont {Y.}~\bibnamefont {Geng}}, \bibinfo {author} {\bibfnamefont {W.}~\bibnamefont {Cui}}, \bibinfo {author} {\bibfnamefont {S.-W.}\ \bibnamefont {Huang}}, \bibinfo {author} {\bibfnamefont {Q.}~\bibnamefont {Zhou}}, \bibinfo {author} {\bibfnamefont {K.}~\bibnamefont {Qiu}},\ and\ \bibinfo {author} {\bibfnamefont {C.}~\bibnamefont {Wei~Wong}},\ }\bibfield  {title} {\enquote {\bibinfo {title} {Soliton bursts and deterministic dissipative {{Kerr}} soliton generation in auxiliary-assisted microcavities},}\ }\href {https://doi.org/10.1038/s41377-019-0161-y} {\bibfield  {journal} {\bibinfo  {journal} {Light: Science \& Applications}\ }\textbf {\bibinfo {volume} {8}},\ \bibinfo {pages} {50} (\bibinfo {year} {2019})}\BibitemShut {NoStop}%
\bibitem [{\citenamefont {Moille}\ \emph {et~al.}(2025{\natexlab{b}})\citenamefont {Moille}, \citenamefont {Shandilya}, \citenamefont {Niang}, \citenamefont {Menyuk}, \citenamefont {Carter},\ and\ \citenamefont {Srinivasan}}]{MoilleNat.Photon.2024a}%
  \BibitemOpen
  \bibfield  {author} {\bibinfo {author} {\bibfnamefont {G.}~\bibnamefont {Moille}}, \bibinfo {author} {\bibfnamefont {P.}~\bibnamefont {Shandilya}}, \bibinfo {author} {\bibfnamefont {A.}~\bibnamefont {Niang}}, \bibinfo {author} {\bibfnamefont {C.}~\bibnamefont {Menyuk}}, \bibinfo {author} {\bibfnamefont {G.}~\bibnamefont {Carter}},\ and\ \bibinfo {author} {\bibfnamefont {K.}~\bibnamefont {Srinivasan}},\ }\bibfield  {title} {\enquote {\bibinfo {title} {Versatile optical frequency division with {{Kerr-induced}} synchronization at tunable microcomb synthetic dispersive waves},}\ }\href {https://doi.org/10.1038/s41566-024-01540-w} {\bibfield  {journal} {\bibinfo  {journal} {Nature Photonics}\ }\textbf {\bibinfo {volume} {19}},\ \bibinfo {pages} {36--43} (\bibinfo {year} {2025}{\natexlab{b}})}\BibitemShut {NoStop}%
\bibitem [{\citenamefont {Moille}\ \emph {et~al.}(2023{\natexlab{b}})\citenamefont {Moille}, \citenamefont {Lu}, \citenamefont {Stone}, \citenamefont {Westly},\ and\ \citenamefont {Srinivasan}}]{MoilleCommunPhys2023}%
  \BibitemOpen
  \bibfield  {author} {\bibinfo {author} {\bibfnamefont {G.}~\bibnamefont {Moille}}, \bibinfo {author} {\bibfnamefont {X.}~\bibnamefont {Lu}}, \bibinfo {author} {\bibfnamefont {J.}~\bibnamefont {Stone}}, \bibinfo {author} {\bibfnamefont {D.}~\bibnamefont {Westly}},\ and\ \bibinfo {author} {\bibfnamefont {K.}~\bibnamefont {Srinivasan}},\ }\bibfield  {title} {\enquote {\bibinfo {title} {Fourier synthesis dispersion engineering of photonic crystal microrings for broadband frequency combs},}\ }\href {https://doi.org/10.1038/s42005-023-01253-6} {\bibfield  {journal} {\bibinfo  {journal} {Communications Physics}\ }\textbf {\bibinfo {volume} {6}},\ \bibinfo {pages} {144} (\bibinfo {year} {2023}{\natexlab{b}})}\BibitemShut {NoStop}%
\end{thebibliography}
%

\end{document}